\makeatletter
\newcommand*\bigcdot{\mathpalette\bigcdot@{.5}}
\newcommand*\bigcdot@[2]{\mathbin{\vcenter{\hbox{\scalebox{#2}{$\m@th#1\bullet$}}}}}
\makeatother

\documentclass[fleqn]{2022AMSE}
\usepackage{graphicx} 
\usepackage{dcolumn} 
\usepackage{dsfont} 
\usepackage{bm} 
\usepackage{tabularx} 
\usepackage{braket} 
\usepackage{siunitx} 
\usepackage{amsmath} 
\usepackage{todonotes} 
\usepackage[ruled]{algorithm2e} 

\DeclareMathOperator{\Div}{div}

\begin{document}

\title{SFFT-based Homogenization:\\
Using Tensor Trains to Enhance FFT-Based
Homogenization}

\author[1, 2]{Sascha H. Hauck}{sascha.hannes.hauck@itwm.fraunhofer.de}%
\author[1]{Matthias Kabel}{matthias.kabel@itwm.fraunhofer.de}
\author[1]{Mazen Ali}{mazen.ali@multiversecomputing.com}
\author[2]{Nicolas R. Gauger}{nicolas.gauger@scicomp.uni-kl.de}

\address[1]{Department of Flow and Material Simulation, Fraunhofer ITWM, Kaiserslautern, 67663 Germany}
\address[2]{Chair for Scientific Computing, University of Kaiserslautern-Landau (RPTU), 67663 Germany}

\abstract{Homogenization is a fundamental technique for estimating the macroscopic properties of materials with microscale heterogeneity.
Among Homogenization methods,
the FFT-based Homogenization algorithm has become widely used due to its computational efficiency and ability to handle complex microstructures.
Nevertheless,
even with GPU acceleration,
FFT-based Homogenization for industrial applications remains excessively time-consuming,
particularly when generating elastic training data for AI models.
This is due to the curse of dimensionality,
which arises from the algorithms reliance on the Fast Fourier Transform,
creating a fundamental bottleneck.

In this paper,
we propose a quantum-inspired SFFT-based Homogenization algorithm that leverages the improved time complexity of a Tensor Train variant of the Quantum Fourier Transform.
By additionally exploiting structural properties of the underlying microstructure,
our method achieves exponential improvements in time complexity and memory efficiency compared to the traditional FFT-based technique - all while remaining executable on classical hardware.
We evaluate the performance of our algorithm across increasingly complex microstructures,
demonstrating its potential advantages and limitations.
}


\setlength{\textheight}{23.6cm}
\thispagestyle{empty}

\maketitle
\setlength{\parindent}{1em}

\vspace{-1mm}
\begin{multicols}{2}

\section{Introduction}
\label{sec:intro}

Homogenization is a mathematical technique used to approximate the macroscopic properties of materials with microscale heterogeneity,
such as composites and
alloys \cite{mei96, jablonski16, wang22, huang23, duan22}.
Computational Homogenization enables simulations of such materials,
even when microstructural features are orders of magnitude smaller than the macroscopic scale \cite{bakhalov89, zhang23, liu25, wang23, gao23}.

One of the most prominent numerical techniques is the FFT-based Homogenization,
widely recognized for its efficiency and accuracy in handling complex microstructures.
Developed by Moulinec and Suquet in 1998 \cite{moulinec98},
this approach offers significant advantages over traditional finite element methods (FEM) \cite{toulemonde08}.
By bypassing the computationally expensive process of assembling local stiffness matrices required in FEM,
the FFT-based method leverages the Fast Fourier Transform (FFT) to iteratively solve the Lippmann-Schwinger equation,
thereby obtaining the macroscopic stiffness tensor \cite{leuschner18, schneider17, schneider21}.

In recent years,
several approaches have sought to improve state-of-the-art FFT-based Homogenization methods.
One such strategy uses GPUs to significantly accelerate the procedure,
however it is severely limited by the memory constrains of the used GPUs \cite{alkhimenkov25}.
Another strategy explores the use of quantum computing,
specifically by utilizing the improved time complexity of the Quantum Fourier Transform (QFT) over the traditional FFT \cite{liu24, givois22}.
While these approaches are promising,
they face significant challenges due to the need for fault-tolerant quantum computing.
As a result,
these methods are unlikely to be feasible in the near future,
particularly within the current Noisy Intermediate-Scale Quantum (NISQ) era.

Another promising approach employs Tensor Networks in a quantum-inspired manner,
aiming to achieve computational speedups through compression.
Although these methods do not face the same challenges as quantum computing,
they still fall short in overcoming the time complexity barrier imposed by the FFT \cite{vondrejc20, risthaus24}.

Recently,
Chen et al. \cite{chen23} demonstrated that the previously unsuccessful attempt to transfer the QFT into a low-rank Tensor Network could be accomplished with minor modifications to the original QFT quantum circuit.
The resulting Tensor Network Operator was an adaptation of the Superfast Fourier Transform (SFFT),
originally introduced by Dolgov et al. in 2012 \cite{dolgov12}.
This adaptation showed significantly improved speed compared to the FFT,
provided the data exhibited favourable structural properties.

\begin{table}[H]
    \caption{List of important symbols and abbreviations used in this work.}
    \label{tab:abbreviations}
    \centering
    \renewcommand{\arraystretch}{1.1} 
    \begin{tabular}{|l|l|}
        \hline
        \textbf{Symbol} & \textbf{Description} \\
        \hline
        TN        & Tensor Network \\
        TT        & Tensor Train \\
        TTO       & Tensor Train Operator \\
        SFFT       & Superfast Fourier Transform \\
        NISQ       & Noisy Intermediate-Scale Quantum\\
        \( n \)    & TT cores / Qubits per Spatial Dimension \\
        \( N \)    & Discretization Points per Spatial Dimension \\
        \( d \)    & Physical Dimension of the TT cores \\
        \( D \)    & Spatial Dimension \\
        \( r \)    & TT rank \\
        \( \delta_\mathrm{tol} \) & Accuracy Threshold for the Algorithms
    \\
        \( \delta_\mathrm{acc} \) & Approximation Error for the TT\\
        \( \delta_\mathrm{trunc} \) & Truncation Threshold for the SVD \\
        \( \varepsilon \) & Strain Field \\
        \( \sigma \) & Stress Field \\
        \( \varGamma_0 \) & Green-Eshelby Operator \\
        \( \mathbf{C}_{(0)} \) & (isotropic) Stiffness Tensor\\
        \hline
    \end{tabular}
\end{table}

In this paper,
we will combine the adaptation of Dolgovs SFFT algorithm with Tensor Network and Homogenization Theory to obtain a novel quantum-inspired SFFT-based Homogenization algorithm.
We will further explore the applicability of our method to various linear elastic geometries in both 2D and 3D. 
A key aspect of our analysis is the memory consumption and speed of our approach,
which we compare to the state-of-the-art FFT-based Homogenization method.
Our goal is to achieve scalability similar to the QFT while ensuring practical computability with classical computers.
Additionally,
our approach has the potential to generate the large amount of training data needed for deep material networks (DMNs) \cite{liu19, gajek20, magino22}. 
In their original formulation,
DMNs are trained exclusively on linear elastic data,
yet they are successfully applied to nonlinear and inelastic problems with remarkable accuracy.

The outline of the remainder of this paper is as follows:
Section \ref{sec:homogen} presents the derivation of the FFT-based Homogenization algorithm and discusses its limitations,
while Section \ref{sec:tensormethod} introduces the Tensor Train format.
Section \ref{sec:qihomogen} presents the 'Zip-Up' algorithm,
which converts our QFT quantum circuit into the desired SFFT Tensor Train Operator,
and concludes with our novel SFFT-based Homogenization algorithm.
In Section \ref{sec:experiments},
we compare the performance of our approach to the state-of-the-art FFT-based method,
focusing primarily on speed and memory consumption.
Before closing,
we provide a short
summary of the paper – along with the statement of the
main conclusions drawn – in Section \ref{sec:conclusion}.
For a comprehensive list of the most important symbols and abbreviations used within this manuscript,
see Table \ref{tab:abbreviations}.

\section{FFT-based Homogenization}
\label{sec:homogen}

The primary objective of the FFT-based Homogenization is to determine the effective linearly elastic macroscopic stiffness tensor $\mathbf{C}^\mathrm{eff}$ of a composite, 
assuming the properties of the individual constituents are known.
The stiffness tensor $\mathbf{C}^\mathrm{eff}$ is defined through the macroscopic stress-strain relationship
\begin{align}
\label{eq:macro_ss}
    \mathbf{\sigma}^\mathrm{eff} = \mathbf{C}^\mathrm{eff} \mathbf{E},
\end{align}
where $\mathbf{\sigma}^\mathrm{eff}$ and $\mathbf{E}$ represent the macroscopic stress and strain tensors, respectively.

To solve the above equation,
it is advantageous to perform the analysis at the local scale of the heterogeneities in the composite.
Within this region,
a representative volume element (RVE) $\mathcal{V}$ is defined,
which is assumed to be significantly smaller than the overall material length scale \cite{cior99}.
This approach,
where a macroscopic problem is addressed by first solving a microscopic problem on an RVE,
is referred to as the Corrector problem \cite{brisard10}.

In this setting,
assuming periodic boundary conditions on the RVE,
the local strain field can be written as
\begin{align}
\label{eq:local_strain}
    \mathbf{\varepsilon(x)}
    = \mathbf{E} + \nabla^\mathrm{S} \mathbf{u(x)},
\end{align}
where $\nabla^\mathrm{S}$ and $\mathbf{u(x)}$ are the symmetrized nabla-operator and the local displacement field,
respectively.
By imposing the condition of a vanishing average
\begin{align*}
    \frac{1}{|\mathcal{V}|}\int_\mathcal{V} \nabla\mathbf{u(x)} \mathrm{d}\mathbf{x}=0,
\end{align*}
it is assured that the macroscopic strain $\mathbf{E}$ represents the mean strain over the RVE.
Since the macroscopic strain is assumed to be known,
the focus of the derivation shifts to finding a representation of the fluctuation term $\nabla^\mathrm{S} \mathbf{u(x)}$.

\begin{table*}[b]
\caption{
Rank Growth and Time Complexity for Different TT Operations (adapted from Kornev et al. \cite{kornev23}).}
\label{tab:time_tt_op}
\begin{threeparttable}
\begin{tabularx}{\textwidth}{|X|X|X|}
\hline
\textbf{Operations} & \textbf{Rank Result} & \textbf{Time Complexity}\\
    \hline
    $z=x*\mathnormal{const}$        & $r(z) = r(x)$        & $\mathcal{O}\left(d r(x)\right)$ \\
    $z=x + y$                       & $r(z) = r(x) + r(y)$ & $\mathcal{O}\left(nd\left[r(x) + r(y)\right]^2\right)$ \\
    $z=x°y$                         & $r(z) = r(x) * r(y)$ & $\mathcal{O}\left(nd r^3(x)r^3(y)\right)$ \\
    $z=Ax$ \textit{(contract)}                & $r(z) = r(x) * r(A)$ & $\mathcal{O}\left(nd r^3(A)r^3(x)\right)$ \\
    $z=Ax$ \textit{(solve z)}               & $r(z)$ & $\mathcal{O}\left(nd r(A)r^3(z)\right)$ \\
    $Az=x$ \textit{(solve z)}                  & $r(z)$               & $\mathcal{O}\left(nd r(A)r^3(z)\right)$ \\
    $z=\mathrm{round}(x, \delta_\mathrm{acc})$ & $r(z) \leq r(x)$     & $\mathcal{O}\left(nd r^3(x)\right)$ \\
    \hline  
\end{tabularx}

\end{threeparttable}
\end{table*}

We continue the derivation by identifying that the local stress tensor must satisfy the force-equilibrium equation,
given by
\begin{align}
\label{eq:force_eq}
    \Div\mathbf{\sigma(x)} 
    &= \Div\left\{\mathbf{C(x)}\left[\mathbf{E} + \nabla^\mathrm{S}\mathbf{u(x)}\right]\right\}    \\
    &= 0.
    \nonumber
\end{align}
To address the local problem, 
equation \eqref{eq:force_eq} is reformulated through an isotropic reference material,
characterized by a constant stiffness tensor $\mathbf{C}_0$ \cite{brisard10},
leading to 
\begin{align}
\label{eq:pol_force_eq}
    \Div\left\{\mathbf{C}_0\left[\mathbf{E} + \nabla^\mathrm{S}\mathbf{u(x)}\right] + \tau(\mathbf{x})\right\} = 0.
\end{align}
with the polarization tensor
\begin{align}
\label{eq:polarization_tensor}
    \tau(\mathbf{x}) = \mathbf{\left[C(x)-C_0\right] \left[E + \nabla^\mathrm{S} u(x)\right]}.
\end{align}
The closed-form differential equation \eqref{eq:pol_force_eq} can be solved using a Green's function $G_0$, 
resulting in the following expression for the displacement gradient
\begin{align}
    \label{eq:green_conv}
    \nabla^\mathrm{S} \mathbf{u(x)} = -\left(\varGamma_0 \star \tau\right)(\mathbf{x}),
\end{align}
where $\varGamma_0= \varepsilon G_0 \Div$ denotes the Green-Eshelby operator for strains \cite{grimm21}.
It is important to note that the divergence operator $\Div$ plays a crucial role in defining the discretization scheme used \cite{willot15},
see Appendix A for a detailed discussion.
By substituting the solution \eqref{eq:green_conv} into the strain expression \eqref{eq:local_strain},
and applying the definition of the polarization tensor \eqref{eq:polarization_tensor},
we obtain the Lippmann-Schwinger equation
\begin{align}
\label{eq:lippmann_schwinger}
       \varepsilon(\mathbf{x}) = \mathbf{E} - \left\{\varGamma_0 \star \left[\left(\mathbf{C - C}_0\right)\varepsilon\right]\right\}(\mathbf{x}).
\end{align}

However,
since the Green-Eshelby operator has a known analytical form in Fourier space,
equation \eqref{eq:green_conv} can be transformed accordingly,
yielding
\begin{align}
    \label{eq:green_conv_momentum}
    \mathcal{F}\left[\nabla \mathbf{u}\right](\mathbf{q}) = -\hat{\varGamma}_0(\mathbf{q})\hat{\tau}(\mathbf{q}), 
\end{align}
where the Fast Fourier Transform $\mathcal{F}[\mathbf{\,\bigcdot\,}]$ is employed for efficient computation.
Throughout this paper,
variables in momentum space will be denoted by a hat symbol to distinguish them from their spatial space counterparts.

Finally,
the FFT-based Homogenization algorithm \ref{algo:fft} is derived by 
incorporating the Fourier representation of the Green-Eshelby operator \eqref{eq:green_conv_momentum} into the Lippmann-Schwinger equation \eqref{eq:lippmann_schwinger}.
The algorithm iteratively refines the initial approximation of the local strain,
conveniently initialized as the mean strain $\mathbf{E}$, by accounting for local variations in the material structure.
 The iterations continue until a predefined accuracy threshold $\delta_\mathrm{tol}$ is achieved.

Working with tensor representations can often be cumbersome,
both in terms of clarity and computational efficiency.
To address this,
we adopt Voigt notation,
which represents stress and strain tensors as 3 (6) dimensional vectors,
while the stiffness tensor is represented as a $3\times3$ ($6\times6$) matrix that operates on them,
within a 2 (3) dimensional space.
This reduction is made possible by exploiting the inherent symmetries of the tensors \cite{schwarz13}.

\begin{algorithm}[H]
    \SetKwInOut{Input}{Input}
    \Input{Mean strain field $\varepsilon_0=\mathbf{E}$}
    \While{not converged}{
        $\tau_m = \left(\mathbf{C} - \mathbf{C}_0\right) \varepsilon_m\;$
        
        $\hat{\tau}_m = \mathcal{F}\left[\tau_m\right]$
        
        $\hat{\varepsilon}_m =\hat{\varGamma}_0 \hat{\tau}_m$
        
        $\varepsilon_{m+1} = \mathbf{E} -\mathcal{F}^{-1}\left[\hat{\varepsilon}_m\right]$
        
        $m = m+1$

        \If{$\|\frac{\varepsilon_{m+1} - \varepsilon_m}{\mathbf{E}} \| < \delta_\mathrm{tol}$}{
            break;
        }
    }
    
    \textbf{Return}: Local strain field $\varepsilon_{m+1}$
    
    \caption{FFT-based Homogenization Algorithm}
    \label{algo:fft}
\end{algorithm}

The effective stiffness tensor $\mathbf{C}^\mathrm{eff}$ can conveniently be determined in Voigt notation,
where the strain-stress relationship \eqref{eq:macro_ss} reduces to a simple matrix-vector product.
By selectively controlling the mean strain $\mathbf{E}$
- where only its $k$-th component is set to unity while all others are zero -
the relationship simplifies to its component form
\begin{align*}
\sigma^\mathrm{eff}_i
= \mathbf{C}^\mathrm{eff}_{ij} \delta_{jk} = \mathbf{C}^\mathrm{eff}_{ik}.
\end{align*}
Using this approach,
the $k$-th column of the effective stiffness matrix can be computed directly from the macroscopic stress vector $\sigma^\mathrm{eff}$.
Consequently, 
by successively setting the mean strain components to unity,
the columns of the effective stiffness matrix can be systematically determined.
Additionally,
the specified mean strains are provided as input to Algorithm \ref{algo:fft}.
The algorithm's output is then used to compute the macroscopic stress vectors $\sigma^\mathrm{eff}$ by averaging over the RVE
\begin{align*}
    \sigma^\mathrm{eff}
    =
    \frac{1}{|\mathcal{V}|}\int_\mathcal{V} \sigma_m(\mathbf{x}) \mathrm{d}\mathbf{x}
    =
    \frac{1}{|\mathcal{V}|}\int_\mathcal{V} \mathbf{C(x)} \varepsilon_m(\mathbf{x}) \mathrm{d}\mathbf{x}.
\end{align*}
This approach allows for the complete computation of the effective stiffness matrix $\mathbf{C}^\mathrm{eff}$.

We conclude this Section by noting that the computational complexity of the FFT-based algorithm \ref{algo:fft} is dominated by the FFT operations,
scaling with $\mathcal{O}\left(D N^D \log N\right)$,
where $N$ is the number of discretization points per dimension $D$.
Thus,
improving the FFT-based Homogenization procedure relies on either enhancing the performance of FFT operations,
such as through hardware acceleration,
or circumventing the classical FFT entirely.

\section{Tensor Train Methods}
\label{sec:tensormethod}

Tensor Networks (TNs) have a rich and diverse history,
having been developed independently in both the mathematics and physics communities \cite{fannes92, hackbusch09, hackbusch15}.
This cross-disciplinary development led to a variety of powerful and complementary approaches across the field.
In the context of TNs,
high-order tensors are decomposed into a network of lower-order tensors,
interconnected through specific contraction patterns.
This allows for efficient manipulation and flexible representation of complex data.

This work specifically focuses on the Tensor Train (TT) format,
along with some of its special cases,
for representing high-order tensors $\mathcal{T}$ \cite{montangere18, biamonte23}.
The elements of $\mathcal{T}$ are related to its TT decomposition as follows 
\begin{align}
\label{eq:tt_format}
    \mathcal{T}(i_1, ..., i_n) = \sum_{\alpha_k} 
    C^{\alpha_1}_{\alpha_0}(i_1)
    C^{\alpha_2}_{\alpha_1}(i_2)
    \, ... \,
    C^{\alpha_n}_{\alpha_{n-1}}(i_n), 
\end{align}
where each physical index $i_k$ runs from $0$ to $d_k-1$.
The novel 3-order tensors $C^{\alpha_k}_{\alpha_{k-1}}(i_k)$ of shape $(r_k\times d_k\times r_{k-1})$
are regarded as the cores of the TT.
The highest dimensionality of all the virtual subspaces is denoted as the rank $r$ of the TT,
providing insight into the feasibility of employing TTs.

The primary advantage of TNs over full tensor representations lies in their data compression capabilities.
An $n$-order tensor,
with each index having a dimension of $d$,
requires exponential memory $O(d^n)$ to store its data.
In contrast, 
the same tensor in the TT format consists of $n$ cores and scales linearly with both its dimension and its number of cores $O(ndr^2)$,
while now additionally depending on its rank.
Thus, 
effective compression is achieved if the rank $r$ remains at a moderately level.

Moreover,
TTs serve not only as a means of compressed storage.
They offer a robust mathematical framework that enables efficient operations,
leading to significant computational speedups when applicable.
The extent of this acceleration depends primarily on the TT rank.
The time complexity and rank evolution of TTs under basic linear algebra are summarized in Table \ref{tab:time_tt_op}.
Given the broad mathematical foundation of TTs,
we refer readers to I. V. Oseledets introduction for a comprehensive discussion for the basic linear algebra required \cite{oseledets11}.

A generalization called the Tensor Train Operator (TTO) can be obtained by extending the TT formula \eqref{eq:tt_format} to include an additional physical index per core
\begin{align}
\label{eq:tto_format}
    \tilde{\mathcal{T}}(\{i_1, j_1\}, ..., \{i_n, j_n\}) = \sum_{\tilde{\alpha}_k}
    &G^{\tilde{\alpha}_1}_{\tilde{\alpha}_0}(i_1, j_1)
    G^{\tilde{\alpha}_1}_{\tilde{\alpha}_0}(i_1, j_1)
    \, ... \,
    \nonumber
    \\
    \times
    &G^{\tilde{\alpha}_n}_{\tilde{\alpha}_{n-1}}(i_n, j_n).
\end{align}
The physical index pairs of each core can be interpreted as incoming and outgoing dimensions,
which allows a TT to be contracted with a TTO.
Therefore,
the TTO serves as a complex transformation on a TT,
and its contraction with a TT is commonly referred to as the 'matrix-vector' product.

A more complex class of operations includes optimization algorithms,
often employed to solve linear systems of equations (LSE).
In this work,
rather than employing such techniques for solving LSEs directly,
we adapt the alternating minimal energy (AMEn) algorithm \cite{dolgov13} to efficiently compute matrix-vector products. 
This adaptation modifies the cost function in the optimization process,
allowing us to obtain an approximate solution $v$ of the equation $Au=\tilde{v}$, with the exact TTs $u$ and $\tilde{v}$ as well as the TTO $A$.
The adaptation specifically minimizes the following cost function
\begin{align*}
    L(\tilde{v}) 
    & = || v - \tilde{v}||^2 \\
    & = \left(v, \tilde{v}\right) - 2 \,\mathfrak{Re}\left(Au, v\right) + const.
\end{align*}
The general derivation closely follows the procedure outlined in the original AMEn paper \cite{dolgov13},
with deviations in relevant details.
Additional information on the adapted method is provided in Appendix B.

The primary advantage of this class of optimization methods over direct contractions between a TTO and a TT becomes evident in the high-rank regime.
The introduced approach not only reduces the overall time complexity but also leads to lower final ranks,
often eliminating the need for an additional rounding step, 
which is discussed further below.
As shown in Table \ref{tab:time_tt_op},
this significantly speeds up subsequent operations.

To efficiently perform linear algebra operations between TTs, converting a regular TT into a TTO can be beneficial.
This is mainly due to the enhanced efficiency of the aforementioned optimization methods.
Assume we have a TT $u$ as defined by equation \eqref{eq:tt_format}.
We can raise it to a TTO $U$,
following the format in equation \eqref{eq:tto_format},
by employing delta functions 
\begin{align}
\label{eq:tt_to_tto}
G^{\alpha_k}_{\alpha_{k-1}}(i_k, j_k)
=
C^{\alpha_k}_{\alpha_{k-1}}(i_k) \delta(i_k, j_k),
\end{align}
where the novel TTO cores $G^{\alpha_k}_{\alpha_{k-1}}(i_k, j_k)$ are 4-order tensors of shape $(d_k \times r_k \times d_k \times r_{k-1})$.
The TTO derived from relation \eqref{eq:tt_to_tto} applied to a TT $v$ produces the same outcome as the Hadamard product of $u$ and $v$.
However,
the rank of the resulting TT exhibits improved scalability,
as it is no longer governed by the product of the ranks of 
$u$ and $v$,
see Table \ref{tab:time_tt_op}.
A detailed proof is provided in Appendix C.

In general, 
operations increase the rank when applied on a TT,
eventually making the format computationally intractable,
as shown in Table \ref{tab:time_tt_op}.
To address this issue,
additional rounding operations must be employed.
Given a tensor $v$ in TT format,
the goal is to find an approximation $\tilde{v}$ such that 
\begin{align}
    \label{eq:rounding_condition_tt}
    ||v - \tilde{v}||_F\leq \delta_\mathrm{acc}||v||_F,
\end{align}
fulfilling the desired accuracy $\delta_\mathrm{acc}$ in terms of the frobenius norm $||\bigcdot||_F$.
The process involves sequential orthogonalization of TT cores followed by truncated Singular Value Decompositions (SVDs), cutting small singular values.
To reach the desired accuracy,
the truncation threshold during the procedure must satisfy the condition
\begin{align*}
    \delta_\mathrm{trunc} = \frac{\delta_\mathrm{acc}}{\sqrt{1-n}}.
\end{align*}
This iterative process preserves essential information while maintaining a reduced rank,
thereby ensuring computational tractability \cite{oseledets11}.

\section{Quantum-Inspired Approach}
\label{sec:qihomogen}
To align the TT format with the mathematical formulation of quantum circuits,
we restrict the dimensionality of the physical indices to 
$d=2$.
This specialized form,
known as the Quantized Tensor Train format,
directly links the dimensionality of the TT cores to the single-qubit subspaces within quantum circuits.
Thus,
an $n$-core TT represents an $n$-qubit quantum state,
with unitary gates acting on these qubits being represented as Quantized Tensor Train Operators.
To improve readability,
we will use the abbreviation TT(O) to additionally refer to its specialized variants.
However,
the specific version in use will always be the one explicitly defined most recently.

The mathematical relationship between the TT format and quantum circuits reveals a closeness in both their mathematical structure and diagrammatic representation.
This duality forms the cornerstone of our quantum-inspired approach.
However,
although the TTOs are derived from unitary operators,
they are generally non-unitary.
This is due to the cut-off of singular values introduced during the TT rounding step to ensure computational tractability.
Further, 
quantum phenomena like entanglement are a special challenge for the TT format, 
since they result in an increased rank and thus an additional computational overhead.
If entanglement grows too large,
Tensor Network based methods in general are likely to reach their computational limits \cite{eisert13}.

One major advantage of quantum-inspired approaches is that they operate entirely on classical computers.
This not only avoids the high error rates of NISQ-era quantum systems but also circumvents the significant challenges associated with transferring classical information to and from quantum computers. \cite{cortese18}.

\begin{figure*}[ht]
\centering
\includegraphics[width=\textwidth]{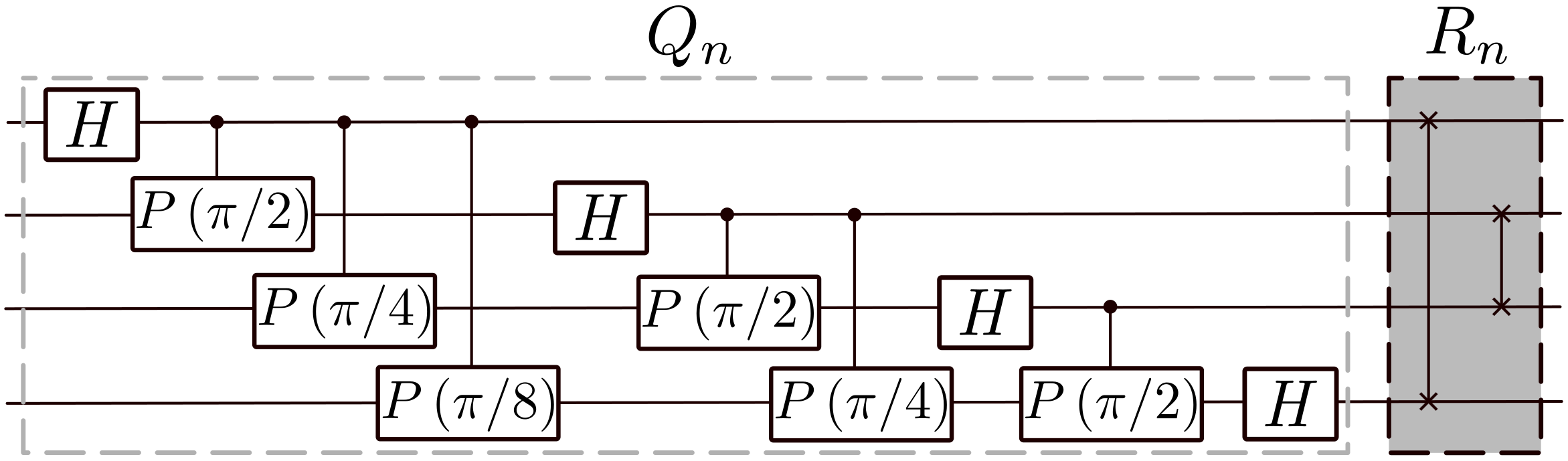}
\caption{Quantum Circuit for the Quantum Fourier Transform $\mathcal{QFT}_n=\mathcal{R}_n \mathcal{Q}_n$ for $n=4$ \cite{nielsen00}. }
\label{fig:qft_qc}
\end{figure*}

\subsection{The Quantum Fourier Transform}
\label{subsec:qft}

The QFT results from a quantum circuit implementation of the classical FFT through the use of Hadamard $H$, (controlled $c$) Phase $P(\theta)$ and $SWAP$ Gates acting on either single or neighbouring qubit pairs
\begin{align*}
    H = \frac{1}{\sqrt{2}}&
    \begin{pmatrix}
        1 & \,\,\,\,1 \\
        1 & -1 \\
    \end{pmatrix},
    \hspace{1cm}
    SWAP = 
    \begin{pmatrix}
        1\,\, & 0\,\, & 0\,\, & 0\,\, \\
        0\,\, & 0\,\, & 1\,\, & 0\,\, \\
        0\,\, & 1\,\, & 0\,\, & 0\,\, \\
        0\,\, & 0\,\, & 0\,\, & 1\,\, \\
    \end{pmatrix},
    \\
    P(\theta) =& 
    \begin{pmatrix}
        1\, & 0 \\
        0\, & \mathrm{e}^{i\theta} \\
    \end{pmatrix},
        \hspace{1cm}
    cP(\theta) = 
    \begin{pmatrix}
        1\,\, & 0\,\, & 0\,\, & 0 \\
        0\,\, & 1\,\, & 0\,\, & 0 \\
        0\,\, & 0\,\, & 1\,\, & 0 \\
        0\,\, & 0\,\, & 0\,\, & \mathrm{e}^{i\theta} \\
    \end{pmatrix}.
\end{align*}
The complete circuit is shown in Figure \ref{fig:qft_qc}.
Alternatively, 
the QFT can be written as a product of two operators $\mathcal{QFT}_n=R_n \,Q_n$.
The $Q_n$ circuit is responsible for the transformation itself, 
while the $R_n$ contribution reorders the output through a succession of $SWAP$ Gates.
The latter one is needed to achieve the same ordering convention as is used within its classical counterpart.

The QFT exhibits an exponentially improved time complexity of $\mathcal{O}\left(D \log^2 N\right)$,
where $N$ is the number of discretization points per dimension $D$,
compared to the FFT.
The latter one scales according to $\mathcal{O}\left(D N^D\log N\right)$,
making the QFT a desirable alternative.

The transfer of the QFT into its corresponding TTO formulation has been done previously,
however,
it was deemed impractical due to its high ranks \cite{dolgov12, garcia21}. 
Recently, Chen et al. \cite{chen23} showed that this may be true for the full QFT,
but not for its reduced form $\mathcal{Q}_n$.
Specifically, they showed that when the reordering operator $\mathcal{R}_n$ is not considered,
the ranks obtained for the TTO formulated reduced QFT decay exponentially with $n$.

This version of the QFT, known as the Superfast Fourier Transform (SFFT) \cite{dolgov12, chen23}, has been observed to significantly outperform the FFT in certain cases, once an input-dependent crossover point is reached.
However,
these cases require the underlying data to be (partially) structured.

We adopt a similar approach by focusing on the reduced form of the SFFT,
which will be derived in the upcoming section.
However,
this results in a different momentum space convention, which will be addressed in the derivation of the SFFT-based Homogenization algorithm.
It is worth noting that on current NISQ-era quantum hardware,
the reordering operation can similarly cause issues due to the numerous error-prone SWAP operators \cite{nowack11}.

\subsection{The Zip-Up Algorithm}
\label{subsec:zip_up}

\begin{figure*}[t]
    \centering
    \includegraphics[width=\linewidth]{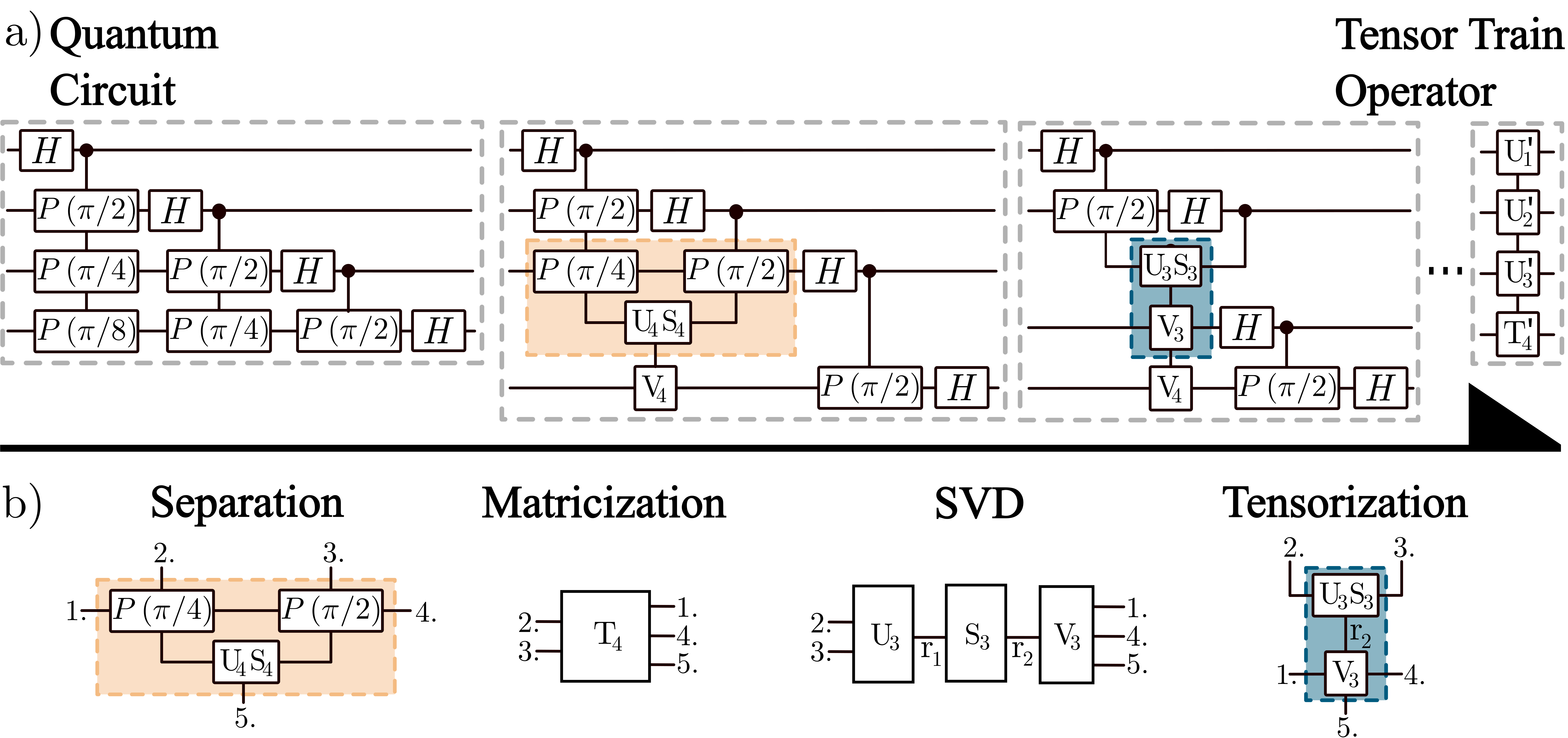}
    \caption{a) The first steps of the Zip-Up algorithm in example of the reduced QFT quantum circuit;
    b) The detailed transformations needed within one step of the Zip-Up algorithm.}
    \label{fig:Zip_Up}
\end{figure*}

The 'Zip-Up' algorithm can be used to convert any quantum circuit into a TTO through a combination of tensor contractions and SVDs \cite{stoudenmire10},
as shown in Figure \ref{fig:Zip_Up}.
In the following,
we will explain the algorithm by transferring the reduced QFT into the SFFT.

\begin{figure}[H]
    \centering
    \includegraphics[width=\linewidth]{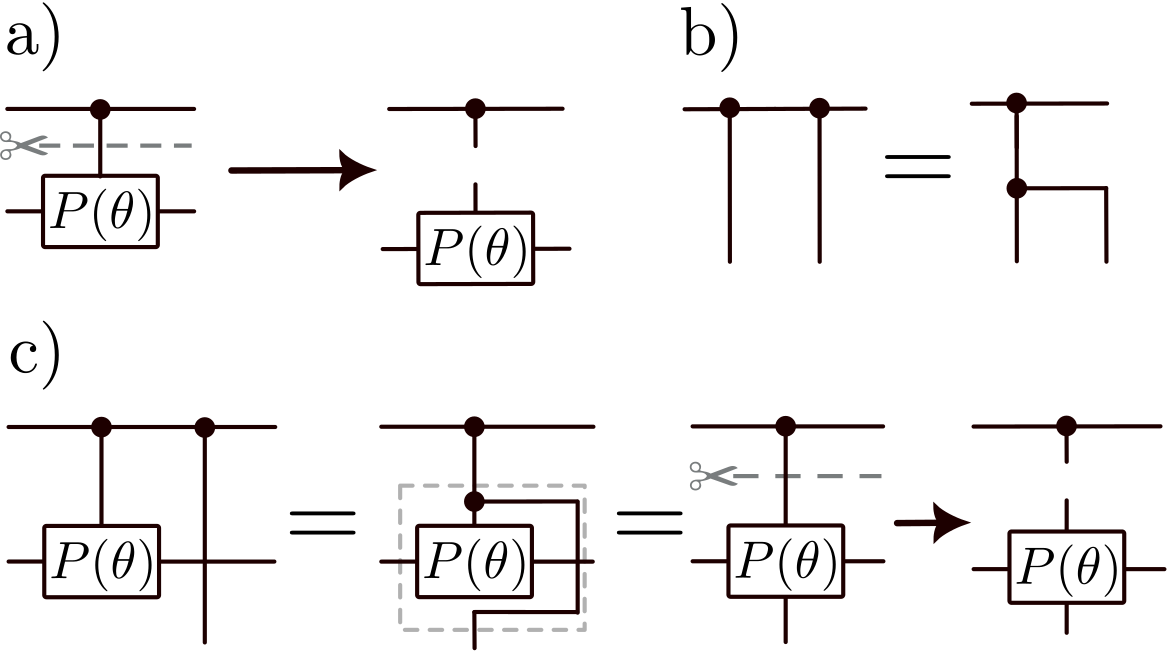}
    \caption{Diagrammatic representation of a) separation, b) reordering and c) recombining of the controlled Phase Gate into Copy, 3-Phase and 4-Phase Gates \cite{chen23}.}
    \label{fig:zip_up_gates}
\end{figure}

Before applying the procedure on the QFT quantum circuit, we must perform preliminary preparations by decomposing the controlled quantum gates and recombining them in a slightly more favourable manner.
In this case,
this means separating the controlled Phase-Gates into a 'control' and a 'phase' component.
Mathematically, this is achieved by representing the full gate as a separated inner product of the following form
\begin{align*}
    cP(\theta) 
    &= \ket{0}\bra{0}\otimes \mathds{1} +\ket{1}\bra{1} \otimes P(\theta) \\
    &= 
    \underbrace{
    \big(\ket{0}\bra{0}\;\; \ket{1}\bra{1}\big) 
    }_{\substack{Copy}}
    \underbrace{
    \begin{pmatrix}
        \mathds{1}\\
        P(\theta)
    \end{pmatrix}.
    }_{\substack{3-Phase}}
\end{align*}
This results in a combination of a Copy Gate and a 3-legged Phase Gate (or simply 3-Phase Gate), 
as shown in  Figure \ref{fig:zip_up_gates} a).
We then separate all controlled Phase Gates into such pairs.
The resulting Copy Gates enforce that all its inputs and outputs correspond to the same state.
Therefore,
the Copy Gates can be interchanged as long as the total incoming and outgoing dimensions,
or 'legs',
remain fixed.
Intuitively,
this can be illustrated through a diagrammatic representation,
as visualized in Figure \ref{fig:zip_up_gates} b). 

These reordered Copy Gates can be recombined with the original 3-Phase Gates,
resulting in novel 4-Phase Gates,
as shown in Figure \ref{fig:zip_up_gates} c).
It should be noted that although we recombine the Copy and 3-Phase Gates,
the incoming and outgoing dimensions have changed. 
With our prepared construct,
we can reformulate the reduced QFT in TTO format using the Zip-Up algorithm.

The algorithm, 
depicted in Figure \ref{fig:Zip_Up} a),
begins by contracting the tensor of the least significant qubit with its right neighbour along their shared index.
In the second step,
this new tensor is reshaped into a matrix,
as shown in Figure \ref{fig:Zip_Up} b).
The incoming and outgoing dimensions of the matrix are determined by the specific leg combinations.
Since the algorithm starts from the bottom and goes to the top,
the vertical as well as the downward legs get combined into the outgoing dimension,
while the upward legs form the incoming dimension.
As a next step,
a SVD is performed on this matrix.
The $V$ matrix remains,
while the $U$ and $S$ matrices get recombined into a new $US$ matrix.
As the final sub-step,
the remaining two matrices are reshaped back into tensor form,
with special care taken to separate the legs back into their original configuration.

In the next iteration of the procedure,
the resulting $V$-tensor remains at the original position, while the $US$-tensor gets contracted with the two tensors acting on the next upper neighbouring qubit.
This procedure is repeated until the most significant qubit is reached. 
Now the procedure restarts by recombining the $V$-tensor acting on the least significant qubit with its right neighbour ($V_4$ in Figure \ref{fig:Zip_Up}).
This down-top process continues until only one tensor acting per qubit is left.

\subsection{SFFT-based Homogenization}
\label{subsec:ttqft_algo}

To derive the SFFT-based Homogenization algorithm, 
the corresponding operands and operators of the FFT-based algorithm \ref{algo:fft} must be transferred into TT and TTO format,
respectively.
The only naturally occurring operators are the FFT and its inverse.
They are simply replaced by the novel SFFT operator $\mathcal{S}$ as well as its inverse ISFFT operator $\mathcal{S}^{-1}$.
The operands - namely,
the local strain $\varepsilon(\mathbf{x})$, 
the mean strain $\mathbf{E}$,
the Green-Eshelby operator $\mathnormal{\hat{\Gamma}}_0(\mathbf{q})$ and the local and isotropic reference stiffnesses $\mathbf{C(x)}$ and $\mathbf{C}_0$ -
must be formulated in TT format. 
To achieve this,
we begin by discretizing the $\mathbf{x}$ or $\mathbf{q}$ dependent operands in a binary fashion.
This means that,
instead of discretizing a function $t(x)$ using a single grid variable $k$,
according to 
\begin{align*}
    t(x) \approx t\left[\left(k-1\right)\Delta x\right] = F(k),
\end{align*}
the different grid points are specified by a multi-index formulation in the form of binary fractions
\begin{align}
\label{eq:binary_discretization}
    t(x) \approx t\left(\frac{b_1}{2} + \frac{b_2}{4} + ... + \frac{b_n}{2^n}\right) = \mathcal{T}(b_1 b_2...b_n).
\end{align}
Although the function is evaluated at the same grid points,
the resulting tensor has a different number of physical indices and corresponding dimensions.
The binary discretization method yields an $n$-order tensor with dimensionality $d=2$ per index, 
which mirrors the physical indices required for our used TT format.
Once the high-order tensor is obtained,
the TT cores can readily be separated by successively applying the (randomized) SVD algorithm or the Tensor-Cross-Approximation algorithm \cite{oseledets11, oseledets10},
as detailed in Appendix D.

In our multidimensional case,
dimensions are processed sequentially,
grouping cores by dimension in its TT representations.
We assume $N=2^n$ discretization points per dimension $D$,
corresponding to $n$ cores (similar to $n$ qubits) per dimension.

Two constant operands are required for the algorithm's iterative procedure:
the mean-strain $\mathbf{E}$ and the isotropic reference stiffness $\mathbf{C}_0$. 
These special cases can be easily represented by a TT of rank one by first generating $(2 \times 2)$ identity matrices and reshaping them into the appropriate core shapes of $(2 \times 1 \times 2 \times 1)$.
Multiplying one of the cores by the constant value of the associated operand results in their respective TT.


Until now,
we have only discussed the internal structures of the tensors in the context of the required discretization.
However, 
the tensors in question generally yield multidimensional outputs, 
introducing additional physical indices.
Here,
we are considering the original tensors in Voigt notation resulting in a simplified operator structure.
One approach to handle these indices is to treat the operands as vectors (or matrices),
with their elements in TT format.
Although this approach is justifiable, 
exploiting the computational benefits of the TT optimization algorithms would not be feasible.
To obtain a full TT representation,
we make a slight deviate from the currently used TT format by adding a single additional core at the beginning of the TT.
This core is permitted to deviate from the imposed dimensionality of $d=2$ and serves as a control for the different components of the Voigt formulation.
Thus,
it holds the information about the Voigt arithmetics rather than the discretization of the function. 
Therefore,
we refer to this version as the controlled Quantized TT.
This approach generalizes naturally to incorporate controlled Quantized TTOs. 
However,
for the sake of readability,
we will not introduce a new abbreviation and will continue to use TT(O) for this specialization.

After obtaining the TT and TTO representations for all operands and operators,
it is computationally beneficial to elevate the TTs involved in the Hadamard multiplications of the algorithm to TTOs,
as discussed in Section \ref{sec:tensormethod}.
This enables efficient contractions or the application of optimization methods,
which are computationally preferable.
This is particularly important for the Green-Eshelby operator,
given its high-rank structure,
for which the adapted AMEn optimization algorithm $\mathcal{A}(\bigcdot, \bigcdot)$ will be employed.

Further,
since the Green-Eshelby operator $\mathnormal{\hat{\Gamma}}_0(\mathbf{q})$ resides in the momentum space, 
it must be treated separately.
This is due to the momentum space convention required to use the reduced QFT as the underlying operator for the SFFT. 
As discussed in Section \ref{subsec:qft}, 
a reordering of the cores is necessary. 
This involves two steps:
1) The indices per geometrical dimension $D$ of the initial discretization tensor,
obtained through equation \eqref{eq:binary_discretization},
must be reversed.
2) The tensor must be complex conjugated.
Performing these additional steps for operators in the momentum space ensures the correct convention according to the SFFT,
yielding our final form of the SFFT-based Homogenization algorithm \ref{algo:sfft}.

Examining Table \ref{tab:time_tt_op},
it becomes clear that the efficiency of the SFFT-based algorithm is highly dependent on the rank dimensions of the operators involved and the strain field.
The total time complexity is given by 
\begin{align}
\label{eq:full_time_complexity}
    \mathcal{O}\Big\{n D \, r_\mathrm{S}^3\left(r_\mathrm{C}^3+ r_\varGamma\right)\Big\}.
\end{align}
with the rank $r_\mathrm{S}$ of the strain TT $\varepsilon_m$,
the rank $r_\mathrm{C}$ of the reduced stiffness TTO $\left[\mathbf{C-C}(\mathbf{x})\right]$ and the rank $r_{\varGamma}$ of the Green-Eshelby TTO $\hat{\varGamma}_0$.
Due to brevity,
we will only state the time complexity of the SFFT-based Homogenization algorithm and provide a detailed derivation in Appendix E.
Our results show that as long as the contributing ranks remain manageable,
an exponential speedup is achieved in both the number of discretization points $N=2^n$ and geometric dimensionality $D$,
in comparison to the FFT-bottlenecked Homogenization method.
The geometry dependent term within the bracket of relation \eqref{eq:full_time_complexity} can be computed during the offline phase,
allowing for an informed estimate on the approaches applicability before running the full algorithm. 

\begin{algorithm}[H]
    \SetKwInOut{Input}{Input}
    \Input{Mean strain field $\varepsilon_0=\mathbf{E}$}
    \While{not converged}{
        $\tau_m = \left(\mathbf{C} - \mathbf{C}_0\right)\varepsilon_m\;$
        
        $\hat{\tau}_m = \mathcal{S}\left[\tau_m\right]$
        
        $\hat{\varepsilon}_m =
\mathcal{A}(\hat{\varGamma}_0,\, \hat{\tau}_m)$
        
        $\varepsilon_{m+1} = \mathbf{E} - \mathcal{S}^{-1}\left[\hat{\varepsilon}_m\right]$
        
        $m = m+1$
        
        \If{$\|\frac{\varepsilon_{m+1} - \varepsilon_m}{\mathbf{E}} \|_F < \delta_\mathrm{tol}$}{
            break;
        }

    }
    
\textbf{Return}: Local strain field $\varepsilon_{m+1}$
    
    \caption{SFFT-based Homogenization Algorithm}
    \label{algo:sfft}
\end{algorithm}

\begin{figure*}[t]
    \centering
    \includegraphics[width=\linewidth]{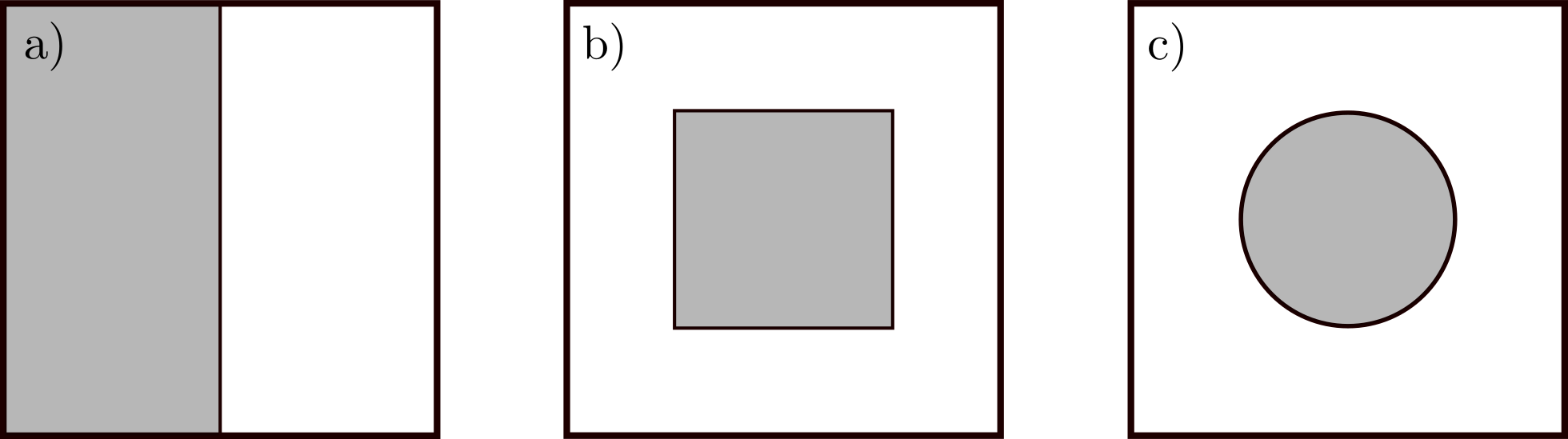}
    \caption{Fundamental 2D geometries used to probe the regimes of the SFFT-based Homogenization algorithm:
    a) Laminate,
    b) Square and
    c) Circle.
    The grey regions have a Young's Modulus of $E_1=29/3\,$\si{\giga\pascal} while the white regions display $E_2=4/3\,$\si{\giga\pascal}.
    The Poisson ratios were set to $\nu_1=\nu_2=1/3$ for the grey and white regions, respectively.}
    \label{fig:geometries}
\end{figure*}

\section{Experiments}
\label{sec:experiments} 

\begin{figure*}[t]
    \centering
    \includegraphics[width=\textwidth]{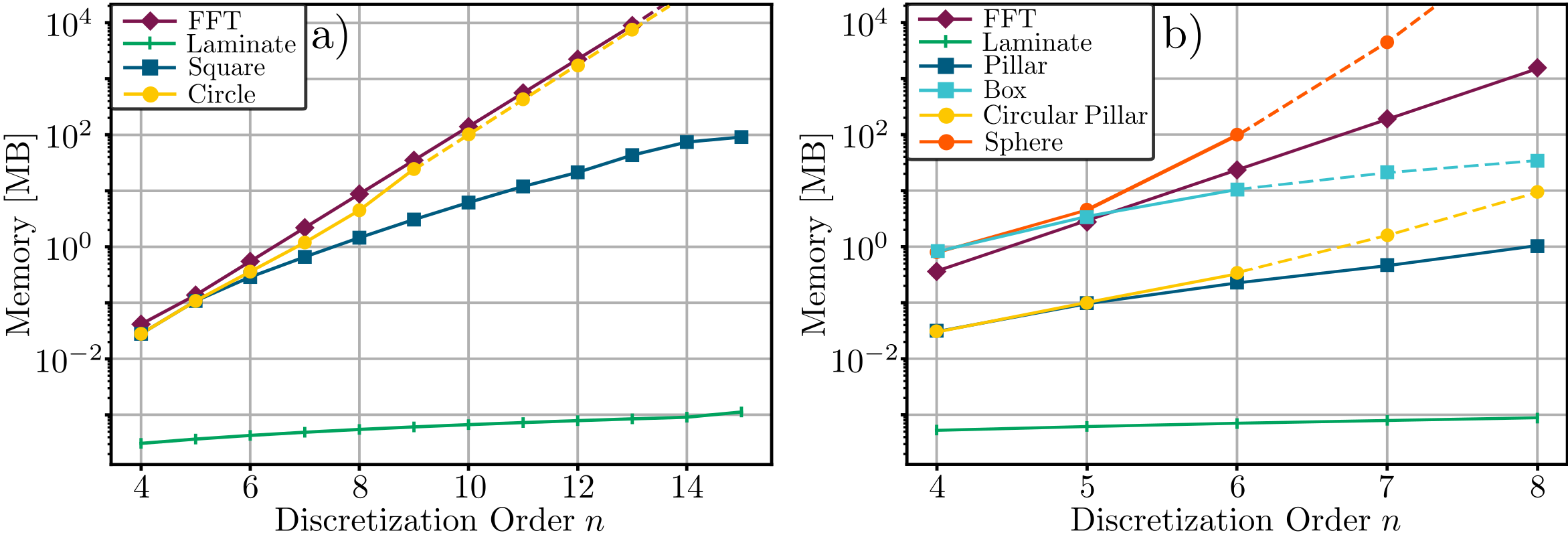}
    \caption{Maximal Memory used for storing the local strain in dependence of the discretization order $n$ in a) 2D and b) 3D space.}
    \label{fig:memory_order}
\end{figure*}

\begin{figure*}[t]
    \centering
    \includegraphics[width=\linewidth]{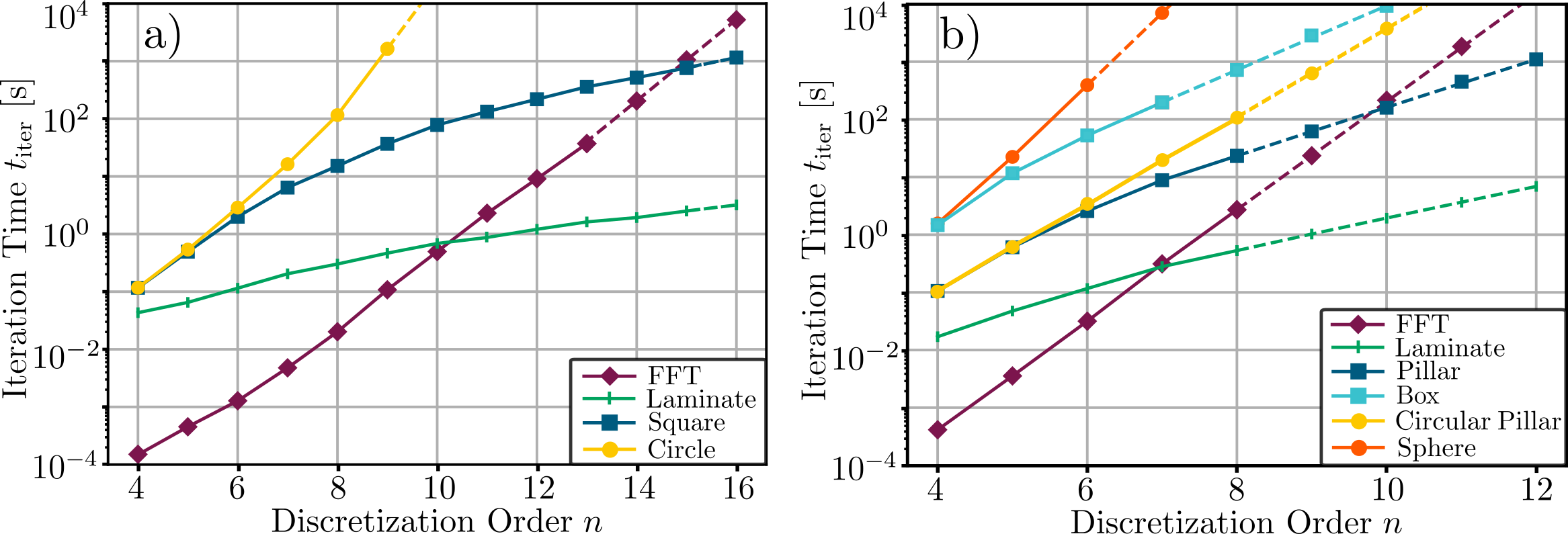}
    \caption{Average iteration time $t_\mathrm{iter}$ in dependence of the discretization order $n$ in a) 2D and b) 3D space.}
    \label{fig:time_order}
\end{figure*}

\begin{figure*}[t]
    \centering
    \includegraphics[width=\linewidth]{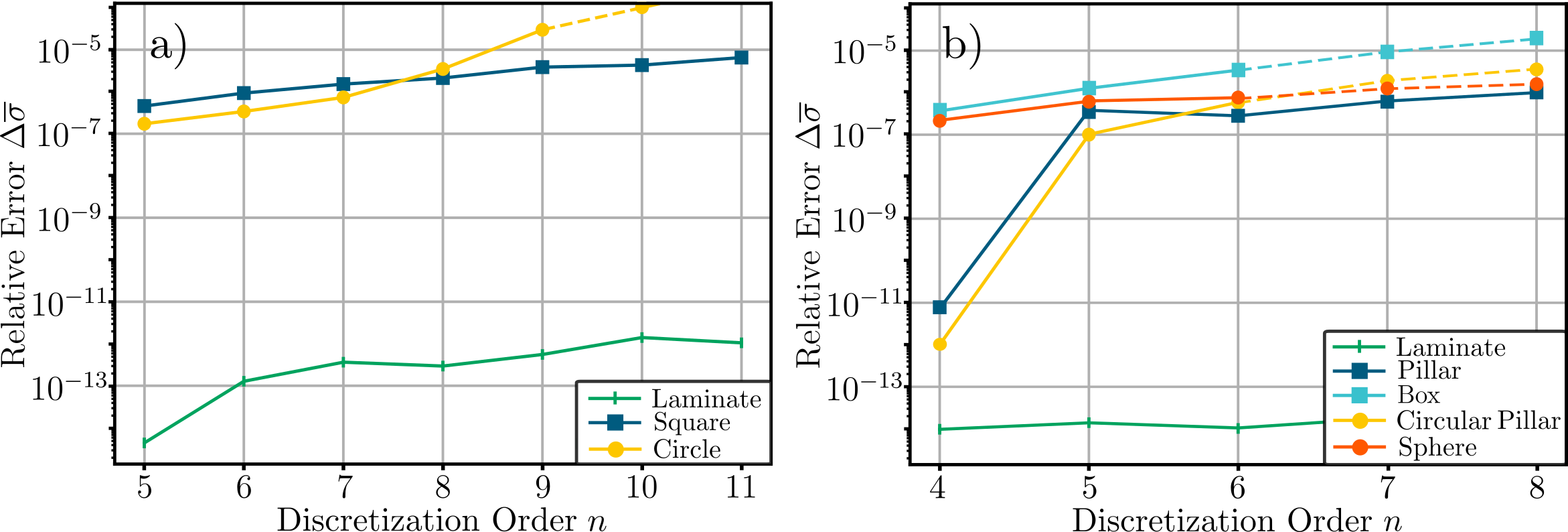}
    
    \caption{Relative Error $\Delta\overline{\sigma}$ in dependence of the discretization order $n$ in a) 2D and b) 3D space.}
    \label{fig:error_order}
\end{figure*}

\begin{figure*}[t]
    \centering
    \includegraphics[width=\linewidth]{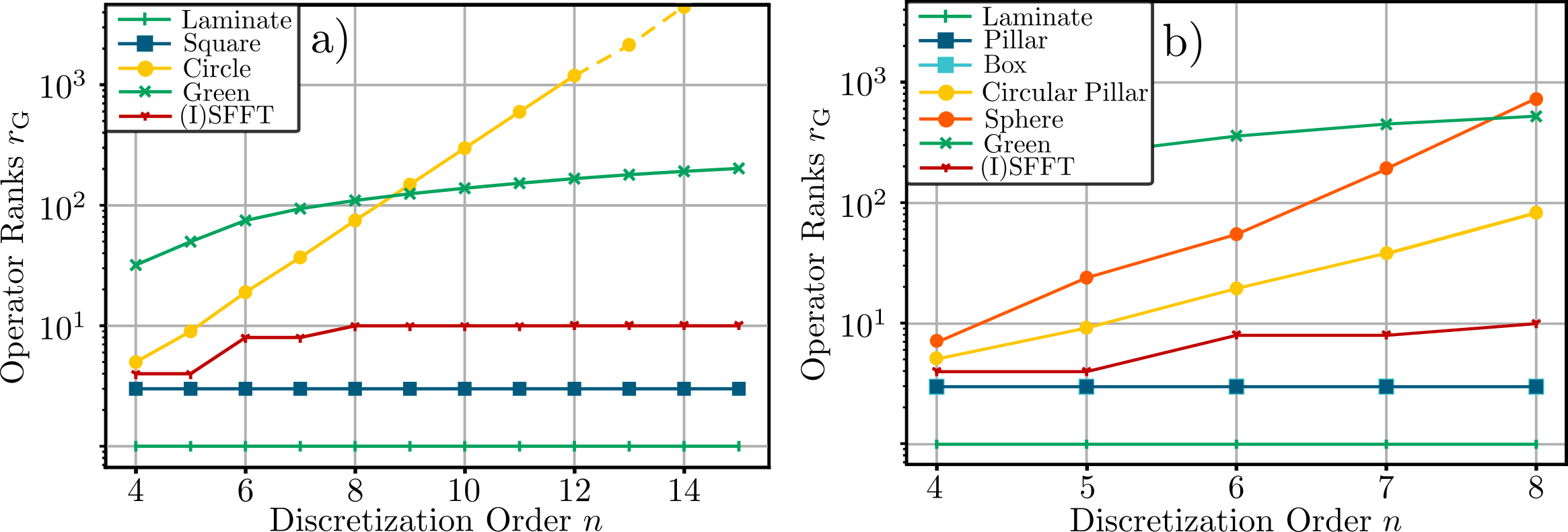}
    \caption{Ranks obtained in dependence of the discretization order $n$: the Stiffness operators $\mathbf{C}$ for the different Geometries,
    the Green-Eshelby as well as the (I)SFFT operators in a) 2D and b) 3D.
    The rounding accuracy was set to $\delta_\mathrm{acc}=10^{-10}$.}
    \label{fig:geometries_rank_order}
\end{figure*}

\begin{figure*}[t]
    \centering
    \includegraphics[width=\linewidth]{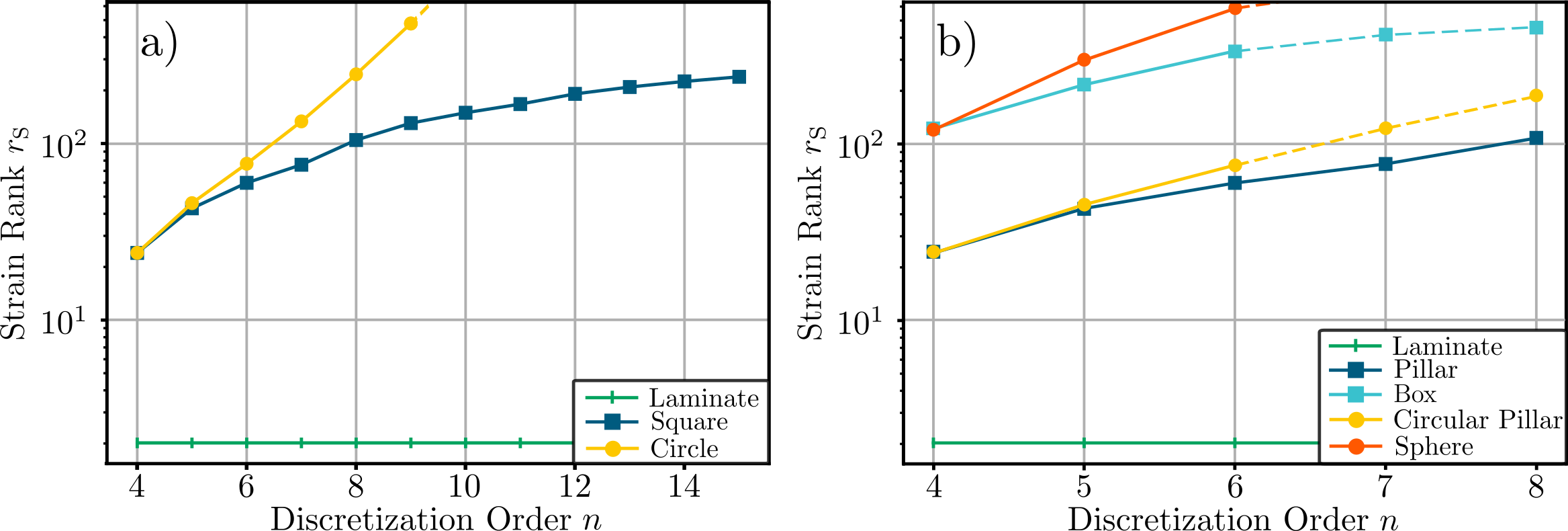}
    \caption{Maximal Strain Rank $r_\mathrm{S}$ during the evaluation of the algorithm in dependence of the discretization order $n$ in a) 2D and b) 3D space.}
    \label{fig:rank_order}
\end{figure*}

\begin{figure*}[!t]
    \centering
    \includegraphics[width=\textwidth]{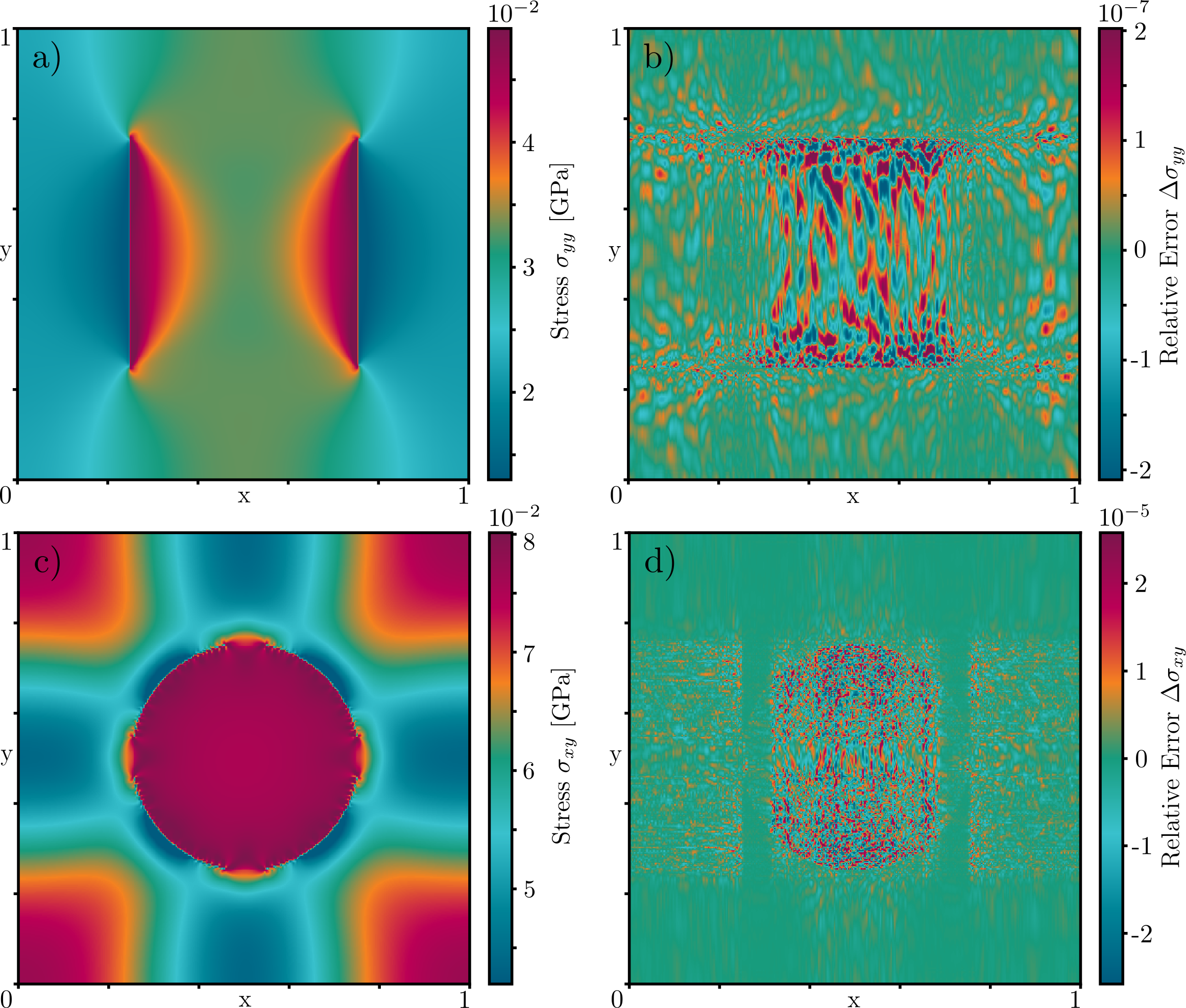}

\caption{a) Stress field $\sigma_{yy}$ under tensile load, and b) point-wise relative error $\Delta\sigma_{yy}$ for the Square geometry.
c) Stress field $\sigma_{xy}$ under shear load, and d) point-wise relative error $\Delta\sigma_{xy}$ for the Square geometry.
The evaluation used $n=8$ for the SFFT- and FFT-based algorithms.}
    \label{fig:density_box}
\end{figure*}

We evaluated the performance of the SFFT-based Homogenization algorithm across various geometries to assess its applicability in diverse scenarios.
The primary geometries used in this study are shown in Figure \ref{fig:geometries}.
Additional tests were conducted on various geometries in both 2D and 3D space,
as detailed in Appendix F.

\subsection{Setup and Preprocessing}
\label{subsec:setup_preprocessing}

Before executing the algorithm,
all operands and operators were converted into the TT format.
The corresponding preprocessing time can be reduced by storing operators in lookup tables for future runs,
as most of them can be reused,
minimizing overhead.
A more detailed exploration of preprocessing scalability and potential improvements can be found in Appendix D.

The geometries shown in Figure \ref{fig:geometries} were analysed in 2D,
with two generalizations considered for 3D.
The first generalization extends the geometries by adding a trivial, 
non-varying z-component,
turning the 2D Laminate into a 3D Laminate,
the Square into a Pillar,
and the Circle into a Circular Pillar.
The second generalization assumes that the geometries in Figure \ref{fig:geometries} are viewed identically from every direction,
resulting in the Square becoming a Box and the Circle becoming a Sphere.
The Laminate is limited to a trivial generalization only.

The accuracy of the SFFT method was evaluated by calculating the relative error in the stress solution field
\begin{align*}
    \Delta\bar{\sigma} = \mathop{\mathbb{E}}
    \left[\frac{|\sigma^\mathrm{SFFT}(\mathbf{x}) - \sigma^\mathrm{FFT}(\mathbf{x})|}{|\sigma^\mathrm{FFT}(\mathbf{x})|}\right],
\end{align*}
where $\sigma^\mathrm{SFFT}$ and $\sigma^\mathrm{FFT}$ are the stress fields obtained after convergence of the SFFT- and FFT-based Homogenization algorithms,
respectively.

Unless otherwise specified,
all experiments used a TT and TTO accuracy parameter of $\delta_\mathrm{acc}=10^{-6}$ and a convergence threshold of $\delta_\mathrm{tol}=10^{-4}$.
The discretization followed a 'staggered grid' approach,
as outlined in Appendix G.
For clarity,
the discretization order $n$ is used throughout,
where $N = 2^n$ represents the number of discretization points per dimension $D$.

Due to memory and time constraints,
not all runs were performed for each discretization order.
In these cases,
missing points are extrapolated and connected by dashed lines in the subsequent graphs.

\subsection{Performance Comparison}
\label{subsecperformance_comparison}

In our experiments, we compared the SFFT-based algorithm \ref{algo:sfft} with the traditional FFT-based approach \ref{algo:fft} by measuring memory consumption and iteration times across different geometries and discretization orders.

The memory consumption results,
shown in Figure \ref{fig:memory_order},
highlight the exponential scaling of the FFT-based approach,
compared to the geometry-specific SFFT method in both 2D and 3D.
For 1D-like geometries,
such as the Laminate,
the SFFT method significantly reduces memory usage.
The Square,
Pillar and Box geometries demonstrate subexponential scaling,
reflecting the adaptability of the SFFT method.
In contrast,
the Circle and
Circular Pillar exhibit exponential scaling similar to the FFT method,
while the Sphere performs even slightly worse than the FFT-based approach.
Notably,
trivially generalized structures,
such as the Pillar and Circular Pillar,
show similar scaling as their 2D counterpart.
This is a direct consequence of the additional (trivial) dimension not notably increasing the runtime for the SFFT-based method due to its geometry-tailored approach.

The iteration times,
shown in Figure \ref{fig:time_order},
reveal that for smaller discretization orders $n$,
the FFT-based method is faster.
However,
our approach outperforms the FFT-based algorithm beyond a geometry-dependent crossover point.
In 2D,
the Laminate crosses over at $n=11$, followed by the Square at $n=15$.
In 3D,
the Laminate reaches the crossover substantially quicker at $n=7$,
while both the Pillar and Circular Pillar geometries reach the crossover at a reasonable order of $n=10$.
Furthermore,
the plot indicates that for even higher discretization orders,
the Box will exhibit a crossover around $n=15$,
aligning with the 2D results.
Although no crossover is expected for the Circle and Sphere,
the Circular Pillar may experience this point at very high discretizations.
While the corresponding discretization order would be far too high for any real-world application,
this may indicate that trivially generalized 3D structures consistently display a crossover point because of their negligible third dimension.
Finally,
it should be noted that the FFT baseline benefits from highly optimized libraries, 
while similarly optimized code is not yet available for the SFFT-based Homogenization algorithm.
Thus,
we anticipate that further optimizations to our implementation will significantly reduce the discretization order at which crossover points occur.

Furthermore,
it is worth noting that the iteration times between the Square and Circle and their trivial extensions
- the Pillar and Circular Pillar -
respectively,
are almost negligible.
The minor difference is primarily due to the non-negligible contribution of the Green-Eshelby operator in 3D.
This mirrors the behaviour observed in the memory scaling experiments.

The relative error $\Delta \overline{\sigma}$ in dependence on the discretization order $n$,
is shown in Figure \ref{fig:error_order}.
For the Laminate,
the analytic solution is achieved after just one iteration, 
resulting in a significantly reduced error.
For all other geometries,
the relative error stabilizes around $10^{-5}$ for both 2D and 3D cases, well within the acceptable range for industrial applications.
Noticeably,
the error remains close to the accuracy of the TT rounding procedure of $\delta_\mathrm{acc}=10^{-6}$,
see equation \eqref{eq:rounding_condition_tt}.

Figure \ref{fig:geometries_rank_order} shows that the SFFT and ISFFT TTO ranks converge to $r_\mathrm{G}=10$ after $n=8$ in both 2D and 3D,
indicating the efficiency and scalability of the respective operators. 
The Green-Eshelby operator exhibits approximately linear rank growth across discretization and dimension,
further demonstrating the favourable scaling of the nongeometry-dependent parts of the SFFT approach.
Therefore,
the applicability of SFFT-based Homogenization is limited by the compressibility of the underlying geometry.
The Circle and its 3D counterparts consistently exhibit exponential rank growth,
highlighting the inherent incompatibility of these geometries with the current SFFT approach.
This is further supported by Figure \ref{fig:rank_order},
where similar growth patterns are observed for the strain rank $r_\mathrm{S}$ of the respective geometries.
In these exponential scaling cases, 
the geometry-dependent term of the time complexity relation \eqref{eq:full_time_complexity} dominates due to the considerable increase of $r_\mathrm{C}$.
This leads to $\mathcal{O}\Big\{n D \, r_\mathrm{S}^3 r_\mathrm{C}^3\Big\}$, 
where both ranks introduce a cubic contribution,
resulting in a rapidly growing compute time. 
Conversely,
in subexponential scaling cases,
the geometry dependent term is typically negligibly,
resulting in an approximate time complexity of $\mathcal{O}\Big\{n D \, r_\mathrm{S}^3 r_\varGamma\Big\}$,
which is primarily governed by the Green-Eshelby TTO and exhibits improved rank scaling behaviour.

Moreover, 
these results suggest that it may be possible to quickly estimate the applicability of our method by compressing the geometries at two different discretization orders.
This approach could provide an indication of the rank and its growth behaviour,
helping to assess whether our method is suitable for a given geometry.

Finally,
it should be noted that in the FFT-based algorithm,
iteration times are primarily limited by the FFT computation itself.
In contrast,
our SFFT-based approach shifts the computational bottleneck to the matrix-vector product between the Green-Eshelby TTO and the strain TT.
This operation typically accounts for over 95\% of the total computation time,
due to the high ranks of both the TT and TTO involved.
This highlights a key area for potential optimization.
As the full scalability limits of the SFFT remain unexplored,
enhancing the efficiency of this matrix-vector product could yield significant performance gains.

\subsection{Solution Fields}
\label{subsec:solution_fields}

The stress solution fields obtained using the SFFT-based Homogenization algorithm were compared to the FFT-based approach for both the Box and Circle geometries,
see Figure \ref{fig:density_box}.
For both shapes, 
the relative error in the solution fields is greatest within the inclusion.
Interestingly,
for the Circle geometry,
a multitude of rectangular regions with similar relative error are observed.
This suggests that the TT format is better suited for handling straight,
non-curved edges in the geometry.
To accommodate curved geometries within the SFFT-based framework,
alternative Tensor Network architectures,
such as projected entangled pair states (PEPS),
the multi-scale entanglement renormalization ansatz (MERA),
or other generalizations of Tree Tensor Networks \cite{shi06, vidal07, orus14},
can be considered.
This approach may be advantageous,
as different Tensor Network formats encode the same structured data in different ways,
leading to varying compression efficiencies \cite{evenbly11}.
However,
this analysis is beyond the scope of the present paper and will be explored in future work.

\section{Conclusion}
\label{sec:conclusion} 

In this paper,
we introduce the SFFT-based Homogenization algorithm,
a geometry-tailored adaptation of the FFT-based Homogenization method that leverages the TT variant of the QFT,
known as the SFFT.
This novel algorithm offers key advantages, 
including improved time complexity and reduced memory consumption,
outperforming the FFT-based approach in scenarios where the geometry can be effectively compressed into a low-rank TT format. 
The advantage stems from the tailored nature of our method,
in contrast to the 'one-size-fits-all' approach of the FFT-based algorithm.
In particular,
memory consumption no longer scales exponentially and consistently outperforms the FFT-based Homogenization algorithm in relevant cases.

While the FFT-based approach retains an advantage for smaller discretizations,
the SFFT-based algorithm surpasses it beyond a geometry-dependent crossover.
Beyond this point,
the improved time complexity of our method leads to a substantial speedup.
These benefits are particularly pronounced in 3D settings with compressible 2D or 1D structures, where the additional dimensions represent redundant data that is processed with high efficiency.
Additionally,
the derived time complexity of our algorithm $\mathcal{O}\Big\{n D \, r_\mathrm{S}^3\left(r_\mathrm{C}^3+ r_\varGamma\right)\Big\}$,
further reinforces the potential for exponential speedup,
provided the contributing ranks remain small.

However,
the current version of our algorithm struggles with geometries featuring circular inclusions,
possibly due to the specific choice of the underlying Tensor Network architecture.
This raises the question of whether alternative network architectures,
such as PEPS,
MERA 
or other Tree Tensor Networks,
could address this shortcoming.

Leveraging GPUs for high discretizations,
made possible by reduced memory consumption,
presents another potential enhancement \cite{pan24}.
This capability is infeasible for traditional FFT-based Homogenization at high discretizations.
Similarly,
utilizing TPUs \cite{ganahl23} and optimizing the underlying software could provide substantial speedups,
potentially offering a decisive advantage to the SFFT-based Homogenization algorithm. 
This is further supported by the observation that the current bottleneck of our algorithm is not due to the theoretical limits of the SFFT,
but rather the optimization method applied to the matrix-vector product between the Green-Eshelby operator $\hat{\varGamma}_0$ and the polarization iterates $\hat{\tau}_m$.
Finally,
we emphasize that our algorithm's results have the potential to significantly accelerate data generation for deep material networks,
enabling the prediction of nonlinear elastic material properties from linear elastic data.

\section*{Acknowledgments}
To enhance readability and ensure comprehensiveness, portions of this manuscript were refined with the assistance of FhGenie.
We thank BMWK for the financial support provided under the EniQmA project. Furthermore, we acknowledge the funding of the German National High Performance Computing (NHR) Association for the Center NHR South-West. 

\bibliographystyle{unsrt}
\bibliography{references}

\end{multicols}

\makeentitle

\renewcommand{\thesection}{Appendix \Alph{section}}  
\setcounter{section}{0} 

\section{Greens operator Discretization Schemes}
\label{app:discretization}

The discretization of computational domains and associated variables is a critical step in numerical modelling,
particularly in Fourier-based methods.
This appendix provides a detailed discussion of the Fourier-space discretization schemes relevant to FFT- and SFFT-based Homogenization algorithms.
These schemes play a central role in ensuring accurate and efficient computations,
particularly by addressing numerical artifacts and ensuring physical consistency in derived quantities.

The specific form of the Green-Eshelby operator determines the choice of the discretization scheme used.
This dependency arises directly from the mathematical formulation of the divergence operator in Fourier space, 
which specifies how variables such as strain and stress are represented and manipulated,
according to
\begin{align*}
    &k_i(\mathbf{q}) \, \sigma_{ij}(\mathbf{q})=0 \\
    &\varepsilon_{ij}(\mathbf{q})=\frac{1}{2}\left[
k_i(\mathbf{q})u_j(\mathbf{q})
    +
k_j(\mathbf{q})u_i(\mathbf{q})
    \right].
\end{align*}

To begin,
the domain $V\equiv L^D$ (here $L=1$) is discretized into $N^D$ pixel/voxels,
resulting in the Fourier modes
\begin{align*}
    q_i = \frac{2\pi}{NL} \xi_i
\end{align*}
with fundamental frequencies
\begin{align*}
    \xi_i = 
    \begin{cases}
        \left(-\frac{N}{2} + 1\right), \,
        \left(-\frac{N}{2} + 2 \right),\,
        \, ... -1, \,0, \,+1, ...
        \left(\frac{N}{2}-1\right),\,\frac{N}{2} 
        &\text{even } N,\\
        \hspace{0.25cm} \,
        \left(-\frac{N-1}{2}\right),\,
        \hspace{0.25cm} \,
        \left(-\frac{N-3}{2}\right),\,
        \, ... -1, \,0, \,+1, ...
        \hspace{0.25cm} \,
        \left(\frac{N-3}{2}\right), \,
        \left(\frac{N-1}{2}\right)
        &\text{odd } N.
    \end{cases}
\end{align*}
These Fourier modes are used to define an explicit representation of the Green-Eshelby operator in momentum space as 
\begin{align}
\label{eq:orig_green_eshelby}
    \hat{\varGamma}^0_{ij,kl}(\mathbf{q}) = \left[q_i\left(q_m \mathbf{C}^0_{mj,kn} q_n\right)^{-1} q_l\right]_\mathrm{sym},
\end{align}
where the subscript '$\mathrm{sym}$' indicates the minor symmetry of the index pairs $(i,j)$ and $(k,l)$, respectively \cite{willot15}.

In continuous space,
the Green-Eshelby operator exhibits the symmetry $\hat{\varGamma}^0(\mathbf{q})^\ast = \hat{\varGamma}^0(\mathbf{-q})$,
which ensures that the physically observable strain 
$\varepsilon$ is real-valued.
For an odd number of discretization points,
this symmetry is inherently preserved in the discretized version of the operator. 
However, 
for an even number of discretization points,
the Nyquist frequency becomes part of the spectrum but lacks a negative counterpart. 
This results in symmetry breaking,
which is a purely numerical artifact that must be addressed accordingly.
The approach we follow enforces symmetry at the Nyquist frequencies $\mathbf{q}_{ny}$ explicitly through
\begin{align*}
    \varGamma^0(\mathbf{q}_{ny}) &\rightarrow \frac{\varGamma^0(\mathbf{q}_{ny})^\ast + \varGamma^0(-\mathbf{q}_{ny})}{2}.
\end{align*}

\begin{table*}[bt]
\caption{Divergence Operator in momentum space for different discretization schemes \cite{willot15, grimm21}.}
\label{tab:discretization_schemes}
\begin{tabularx}{\textwidth}{|X|X|}
\hline
\bf{Scheme Name} & \bf{Divergence Operator $k_j(\mathbf{q})$}\\
    \hline
    Direct    & $k_j^D(\mathbf{q}) = \mathrm{i} q_j$ \\
    Centered  & $k_j^C(\mathbf{q}) = \mathrm{i} \sin\left(q_j\right)$ \\
    Forward    & $k_j^F(\mathbf{q}) = +\left(\mathrm{e}^{+\mathrm{i}q_j} - 1\right)$ \\
    Backward    & $k_j^F(\mathbf{q}) = -\left(\mathrm{e}^{-\mathrm{i}q_j} - 1\right)$ \\
    Hex8R     &  $k_j^H(\mathbf{q}) = \frac{\mathrm{i}}{2}\tan\left(\frac{q_j}{2}\right)\sum_l\left(1 + \mathrm{e}^{\mathrm{i}q_l}\right)$  \\
    Staggered & $k_j^\pm(\mathbf{q}) = \pm\left(\mathrm{e}^{\pm\mathrm{i}q_j} - 1\right)$\\
\hline
\end{tabularx}
\end{table*}

By combining the Green-Eshelby operators derived from the different discretization schemes listed in Table \ref{tab:discretization_schemes} -excluding the staggered grid scheme -with equation \eqref{eq:orig_green_eshelby},
we obtain the compact form
\begin{align*}
    \hat{\varGamma}^{0,\mathrm{gen}}_{ij,lm}\left(\mathbf{k}\right)
    =
    \frac{
    \Big(\lambda^0 + 2\mu^0\Big)
    \Big(k_i k_j\delta_{jl}\Big)_\mathrm{sym}
    +
    \lambda^0
    \Big(k_i k_m^\ast s_{ijl}\Big)_\mathrm{sym}
    }{
    \mu^0\left[ 2\left( \lambda^0+\mu^0\right) - \lambda^0 |\mathbf{k}^2|^2\right]
    }
    -
    \frac{
    \Big[ 
    \frac{\lambda^0}{\mu^0}
    \Re\left(k_i k_j^\ast\right)
    \Re\left(k_l k_m^\ast\right)
    +   
    k_i k_j k_l^\ast k_m^\ast
    \Big]
    }{
    |\mathbf{k}|^2\left[ 2\left( \lambda^0+\mu^0\right) - \lambda^0 |\mathbf{k}^2|^2\right]
    }.
\end{align*}
where $\lambda^0$ and $\mu^0$ are the lamé parameters of the underlying geometry \cite{willot15}.
Additionally, 
the symmetry operator $s_{ijk}$ was introduced
\begin{align*}
    s_{ijl} = 
    \begin{cases}
        +\frac{4}{|\mathbf{k}|^4} \mathfrak{Im} \left(k_i k_k^\ast\right)^2
        & i\neq j=l,\\
        -\frac{4}{|\mathbf{k}|^4} \mathfrak{Im} \left(k_l k_j^\ast\right) \mathfrak{Im} \left(k_l k_i^\ast\right)
        & i \neq j \neq l \neq i,\\
        0 & \mathrm{else}.
    \end{cases}
\end{align*}
The staggered grid discretization scheme differs from the generic schemes due to its dual formulation,
which simultaneously employs both backward and forward discretization schemes.
For this case,
the Green-Eshelby operator \cite{grimm21},
see equation \eqref{eq:orig_green_eshelby},
can be reformulated as
\begin{align*}   \hat{\varGamma}^{0,\mathrm{st}}_{ij,lm}\left(\mathbf{k}^\pm\right)
    =
    -\frac{1}{4\mu^0 \left|\left(\mathbf{k}^\pm\right)^4\right|}
    \Bigg\{
    \left[\mathbf{k}^-_j + \delta_{ij}\left(\mathbf{k}^+_j - \mathbf{k}^-_j \right)\right]
    \left[\mathbf{k}^+_m + \delta_{lm}\left(\mathbf{k}^-_m - \mathbf{k}^+_m \right) \right]
    \left(\delta_{il}
    +\frac{\lambda^0 + \mu^0}{\lambda^0 + 2\mu^0}
    \mathbf{k}^-_i\mathbf{k}^+_l\right)
    \Bigg\}_\text{sym}.
\end{align*}

\section{Approximate Tensor Train Arithmetic}
\label{app:approx_TT_arithmetic}

A significant class between TTs and TTOs consist of optimization algorithms to solve systems of linear equations or to obtain an approximate solution to a matrix-vector product. 
There are many different approaches that were developed to solve these problems. 
The most famous ones are the alternating least square (ALS) method \cite{holtz12},
the density matrix renormalization group (DMRG) algorithm \cite{white93}
and the alternating minimal energy (AMEn) algorithm \cite{dolgov13}.
The latter one being regarded as the current state-of-the-art method.
Here we will use an adaptation of the classical AMEn algorithm to solve the matrix-vector product of a TT with a TTO.

We are interested in finding an approximate solution $u$ to the equation $Au=\tilde{v}$,
with the exact TTs $u$ and $\tilde{v}$.
Thus, 
our goal is to minimise the following cost function
\begin{align*}
    L(\tilde{y}) 
    & = || v - \tilde{v}||^2 \\
    & = \left(v, v\right) - 2 \mathfrak{Re}\left(Au, v\right) + const.
\end{align*}
This high-dimensional optimisation problem can be reduced to a local-optimisation over the different 
\begin{align*}
    L(\tilde{y}) 
    & = \left(y_k, y_k\right) - 2 \mathfrak{Re}\left(\mathcal{Y}^\ast_{\neq k} A \mathcal{X}^\ast_{\neq k} x_k, y_k\right) + const. 
\end{align*}
In the above equation we used the projection operators defined through 
\begin{align*}
    y = \mathcal{Y}^\ast_{\neq k} y^{(k)},
    x = \mathcal{X}^\ast_{\neq k} x^{(k)},
\end{align*}
where $y_k$ and $x_k$ are the $k$-th core of $y$ and $x$, respectively \cite{dolgov13}.

It can readily be verified that the gradient of the cost function is zero when the corresponding Galerkin condition is met
\begin{align}
    \label{eq:galerkin_condition}
    \left(\mathcal{Y}^\ast_{\neq k} A \mathcal{X}_{\neq k}\right) x_k = y_k.
\end{align}

First, an initial guess for the solution is provided.
The second step involves updating the first core of $y$,
denoted as $y_1$,
by minimizing the cost function locally. This corresponds to solving the Galerkin condition given in equation \eqref{eq:galerkin_condition}.

In the third step, the residual $z=\tilde{y}-Ax$ is computed. This step is generally computationally expensive and thus undesirable; therefore, an approximate residual $\tilde{z}\approx z$ is used instead. Once the approximate residual is obtained, its first core is used to enrich the original first core of the solution $y_1$.
This step, known as the Galerkin correction, not only allows for incremental refinement but, more importantly, enables an adaptive rank procedure.
It is important to note that rank adaptation is not exclusive to the AMEn algorithm; similar approaches can be found in other solvers as well.

Finally, the first core undergoes orthogonalization via QR-decomposition.

This procedure is recursively applied to all cores,
and thus to all local optimization problems.

The main advantage of this class of optimization methods, compared to an exact solution, lies not only in its improved time complexity but also in the resulting rank.
If the result $y$ has a rank similar to the input TT $x$, the time complexity improves by a factor of $r(A)^2$,
Moreover,
the rounding procedure becomes noticeably faster,
as it depends on the cubic rank.
With a quasi-exact solution, the final rank scales with the product of the matrix and vector ranks.
Finally, we want to note that we have used an improved version of the TTO-TT matrix-vector product built on the matrix-vector within the TT-Toolbox available on GitHub.

\section{Connection between a raised Tensor Train and its Hadamard Product}
\label{app:proof_TTO-TT}

Lets assume we have two TTs $a$ and $b$, 
defined as follows
\begin{align*}
a(i_1,i_2, ...i_n) &= a(i_1) a(i_2) ... a(i_n), \\
b(i_1,i_2, ...i_n) &= b(i_1) b(i_2) ... b(i_n).
\end{align*}
The next step is to raise the TT $a$ to a TTO $A$,
by defining the TTO cores as
\begin{align}
\label{eq:TTO_raise}
    A(i_k, j_k)= a(i_k) \delta(i_k, j_k).
\end{align}
where $\delta(\cdot, \cdot)$ is the Kronecker delta function. 
This operation effectively "raises" the TT into a higher-dimensional operator form.
Thus, the TTO $A$ can be written as
\begin{align*}
    A(\{i_1, j_1\}, \{i_2, j_2\},... \{i_n, j_n\}) 
    &= A(i_1, j_1) A(i_2, j_2)... A(i_n, j_n)\\
    &= a(i_1)\delta(i_1, j_1) a(i_2)\delta(i_2, j_2)... a(i_n)\delta(i_n, j_n).
\end{align*}
Next,
we will show that the contraction of the TTO $A$ with another TT $b$ results in a new TT $c$,
which is the Hadamard product of the two TTs $a$ and $b$.
The contraction of $A$ and $b$ involves summing over the intermediate indices $j_k$ 
\begin{align*}
   c(i_1, i_2, ... i_n) 
   &= \sum_{j_k} A(\{i_1, j_1\}, \{i_2, j_2\},... \{i_n, j_n\}) b(j_1,j_2, ...j_n)\\
   &= \sum_{j_k} A(i_1, j_1) A(i_2, j_2)... A(i_n, j_n) b(j_1) b(j_2)...b(j_n) \\
   &= \sum_{j_k} A(i_1, j_1) A(i_2, j_2)... A(i_n, j_n) b(j_1) b(j_2) ...b(j_n)\\
   &= \sum_{j_k} a(i_1)\delta(i_1, j_1) a(i_2) \delta(i_2, j_2)... a(i_n) \delta(i_n, j_n) b(j_1) b(j_2) ...b(j_n)\\
   &= \big[a(i_1) a(i_2) ... a(i_n)\big]\, \big[b(i_1) b(i_2) ...b(i_n)\big]\\
   &= a(i_1, i_2,... i_n)\, b(i_1, i_2, ...i_n),
\end{align*}
where we used the identity from equation \eqref{eq:TTO_raise} in the fourth line to simplify the expression.
The Kronecker delta functions ensure that the contraction of the TTO with the TT results in a component-wise product between the two tensors.

Thus,
raising a TT $a$ to a TTO and contracting it with another TT $b$ is equivalent to performing the Hadamard product between the two original TTs $a$ and $b$.
This shows that the contraction procedure preserves the structure of the TT and transforms the operation into a simple element-wise multiplication,
which is computationally efficient, 
enables the usage of optimization methods and is easy to handle in practice.

\section{Scalable Implementation}
\label{app:implementation}

The implementation of the SFFT-based Homogenization algorithm needed to achieve a certain scalability threshold for high-resolution scenarios.
In this case,
characterized by a large number of discretization points per dimension,
we observed significant memory and time consumption during the SFFT-based algorithm's preparation phase.
This issue arises due to the reliance on the state-of-the-art TT-SVD algorithm \cite{oseledets11},
which is computationally constrained by the usage of the SVD.

As such, the SVD step scales with a complexity of $\mathbf{O}\left(mn^2\right)$ for an $m \times n$ matrix,
where $n<m$.
Since higher-resolution matrices scale exponentially in size,
this results in exponential time consumption for separating the cores in the TT or TTO representations.
Such behaviour is highly undesirable,
as it merely shifts the computational burden from the iterative phase to the preprocessing phase.
Thus,
instead of using the TT-SVD method,
we will use a combination of the streamable tensor-train approximation (STTA) \cite{kressner23} and an updated TT-SVD method build on the randomized SVD.
The choice between these methods depends on the computational and memory restrictions of the compute system used.

The STTA approach can be used in a multi-core setting using pythons dask library \cite{karau23} and is especially useful for matrices with a lower and intermediate resolution. 
In the high-resolution regime, we are typically encountering too high memory restrictions for the multi-core system to be beneficial. 
This is either due to the system not having enough memory per core or due to the high overhead of transferring the compute-graph between the nodes.
In this case, the slower non-multicore version of the randomized TT-SVD with additional rank truncation threshold can be used.
While the SVD of a rank $k$ matrix with shape $m\times n$ has a time complexity $\mathcal{O}\left(mn *\min(m, n)\right)$ that scales quadratic, the randomized SVD only scales linear in the matrix dimensions according to $\mathcal{O}\left(mn * \log(k) + \left(m+n\right)*k^2\right)$ \cite{halko11}.
Thus, we are able to speed up the process by at least a quadratic factor, which represents a significant improvement for large matrices. Alternatively, the TT-cross approximation can be used.
In this case,
a cross-approximation strategy is employed,
where the most important components of the tensor are iteratively selected to provide the best reduced rank approximation.
TT-cross selects a small subset of the tensor's entries for its approximation, which efficiently captures the tensors structure \cite{oseledets10}. 
However,
since (randomized) TT-SVD is more accurate,
we opted not to use TT-cross for obtaining the optimal solutions for our derived TT.

To improve scalability,
we need to consider the case of reusing preprocessed computations for subsequent runs.
Almost all TTs and TTOs within the SFFT-based Homogenization algorithm depend only on the dimensionality, the chosen resolution, and the core ordering. This includes both the (inverse) SFFT and the initial strain TT. As a result, these components can be saved in a lookup table for later use across multiple problems that share a common set of system parameters.
A special case in this context is the geometry-dependent stiffness operator
$\mathbf{C}(\mathbf{x})$ as well the Green-Eshelby operator $\hat{\varGamma}_0(\mathbf{k})$.
The latter depends on the chosen Lamé parameters as well as the discretization scheme used.
Since the Green-Eshelby operator inherently defines the discretization choice for the problem, recomputation is generally unavoidable for different schemes. However, this is not necessary for varying sets of Lamé parameters $\{\mu, \lambda\}$.
In such cases, we can reuse the Green-Eshelby operator, which can generally be expressed in the following form
\begin{align}
    \label{eq:green_separability}
    \hat{\varGamma}_0(\mathbf{k}) = \sum_{i}\alpha_i(\mu, \lambda) \hat{\varGamma}^{(i)}(\mathbf{k}).
\end{align}
Therefore, the reduced Gamma functions $\hat{\varGamma}^{(i)}(\mathbf{k})$ can be stored separately in TT format for efficient reuse. When needed, the prefactors $\alpha_i(\mu, \lambda)$ can easily be computed and the saved reduced Gamma functions can be recombined according to equation \eqref{eq:green_separability}.

Finally, in contrast to all other quantities in the SFFT-based Homogenization algorithm, the stiffness operator $\mathbf{C}(\mathbf{x})$ must be recomputed each time a new geometry is used.

\section{Time Complexity}
\label{app:timecomplexity}

For the following discussion,
we distinguish between the following ranks
\begin{itemize}
    \item $r_\mathrm{S}$ for the strain TT,
    \item $r_\mathrm{\tau}$ for the strain TT in momentum space,
    \item $r_\mathrm{C}$ for the raised reduced stiffness TTO $\left[\mathbf{C}-\mathbf{C}(\mathbf{x})\right]$,
    \item $r_\mathrm{F}^{(i)}$ for the (inverse) SFFT TTO $\mathcal{S}^{(-1)}$,
    \item $r_\varGamma$ for the raised Green-Eshelby Operator TTO $\hat{\varGamma}_0$.
\end{itemize}
While $r_\mathrm{F}^{(i)}$ is determined solely by the chosen resolution, 
$r_\mathrm{\varGamma}$ also depends on the discretization scheme used.
The rank $r_\mathrm{C}$ is geometry dependent and grows with increasing structural complexity.
The strain rank $r_\mathrm{S}$ is influenced by all previously mentioned factors.
We assume that during the initial few iterations,
$r_\mathrm{S}$ grows until it reaches an approximate saturation point,
after which it remains stable in the spatial domain. 
The rank can still change slightly during the iterative process, 
however, 
these changes are negligibly small.
In momentum space,
we make a similar assumption,
but instead of matching the spatial rank,
we assume it follows a linear relation
$r_\mathrm{\tau}= a r_\mathrm{S}$,
with some small transfer constant $a$.
Finally,
the mean-strain TT exhibits a trivial rank of 1 and is thus negligible.
Using these assumptions on algorithm 2 
and the data from Table 2, 
we can readily derive the time complexity of our novel SFFT-based Homogenization algorithm as
\begin{align}
\label{eq:full_time_complexity}
    \mathcal{O}\Big\{n D \, r_\mathrm{S}^3\left[\left(1+a^3\right)r_\mathrm {f}^3+r_\mathrm{C}^3+a r_\varGamma\right]\Big\}.
\end{align}
We assume that the SFFT and its inverse have approximately the same rank,
as validated in Section 5.
We also use the geometric dimensionality $D$ instead of the core dimensionality $d$,
since the largest core dimension determines the time complexity.
In our quantized TT approach,
the first core,
governed by the Voigt formalism,
sets the largest core dimension.
For the small $D$ regime,
this leads to an approximately linear relationship between 
$D$ and the Voigt dimensionality of the first core,
with an additional multiplicative offset.
The dominant factor in computational complexity is the strain rank $r_\mathrm{S}$,
influenced by the solution field smoothness and geometry.
We simplify the time complexity derived in \eqref{eq:full_time_complexity} using empirical data from Section 5.
First,
the transfer constant satisfies 
$a \approx 1$.
Second,
the significantly lower rank of the (inverse) SFFT TTO compared to the strain-field and Green-Eshelby operator shows that the SFFT contribution is negligible small.
This results in the simplified time complexity of algorithm 2 given by
\begin{align}
\label{eq:reduced_time_complexity}
    \mathcal{O}\Big\{n D \, r_\mathrm{S}^3\left(r_\mathrm{C}^3+ r_\varGamma\right)\Big\}.
\end{align}

The above reduced time complexity shows a significant dependency on the geometry since both the ranks $r_\mathrm{S}$ and $r_\mathrm{C}$ depend on it. 
Furthermore,
since both ranks have a cubic contribution,
it indicates that only for geometries with low rank
the asymptotic scaling of our novel algorithm is beneficial.

\begin{figure}[H]
    \centering
    \includegraphics[width=0.85\textwidth]{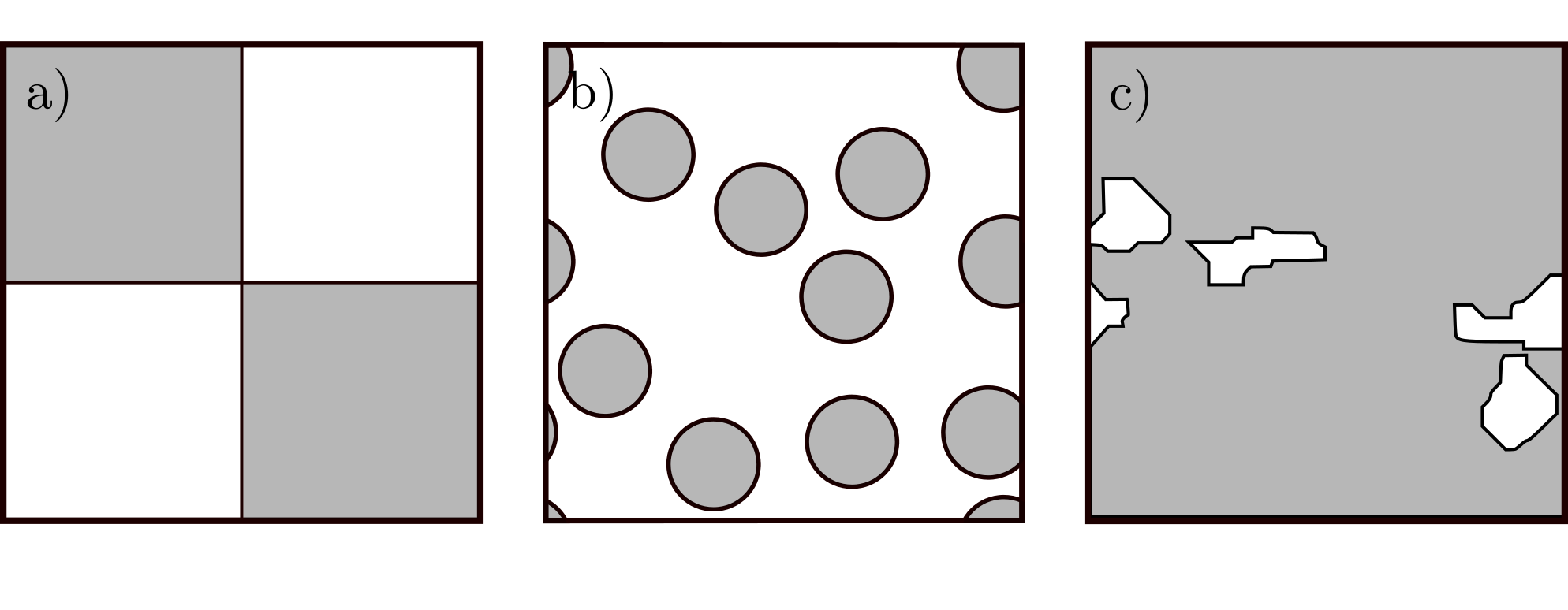}
    \caption{Additional 2D geometries used to probe the regimes of the SFFT-based Homogenization algorithm:
    a) Checkerboard,
    b) Spots and
    c) Voronoi.
    The grey regions have a Young's Modulus of $E_1=29/3\,$\si{\giga\pascal} while the white regions display $E_2=4/3\,$\si{\giga\pascal}.
    The Poisson ratios were set to $\nu_1=\nu_2=1/3$ for the grey and white regions,
    respectively.}
    \label{fig:appendix_geometries}
\end{figure}

\section{Additional Geometries}
\label{app:additional_geometries}

We conducted a performance comparison between the SFFT-based algorithm and the traditional FFT-based method, evaluating memory usage and iteration times across various geometries and discretization orders.

The geometries in Figure \ref{fig:appendix_geometries} were examined in 2D, with two possible generalizations extended to 3D space. The first generalization adds a simple, non-varying z-component, transforming the Spots geometry into Fibres and the Checkerboard into the reduced Checkerboard. The second generalization assumes that the geometries in Figure \ref{fig:appendix_geometries} are viewed identically from all directions, resulting in the Checkerboard geometry becoming Checkerboard 3D and the Voronoi geometry becoming Voronoi 3D.

\begin{figure}[ht]
    \centering
    \includegraphics[width=\textwidth]{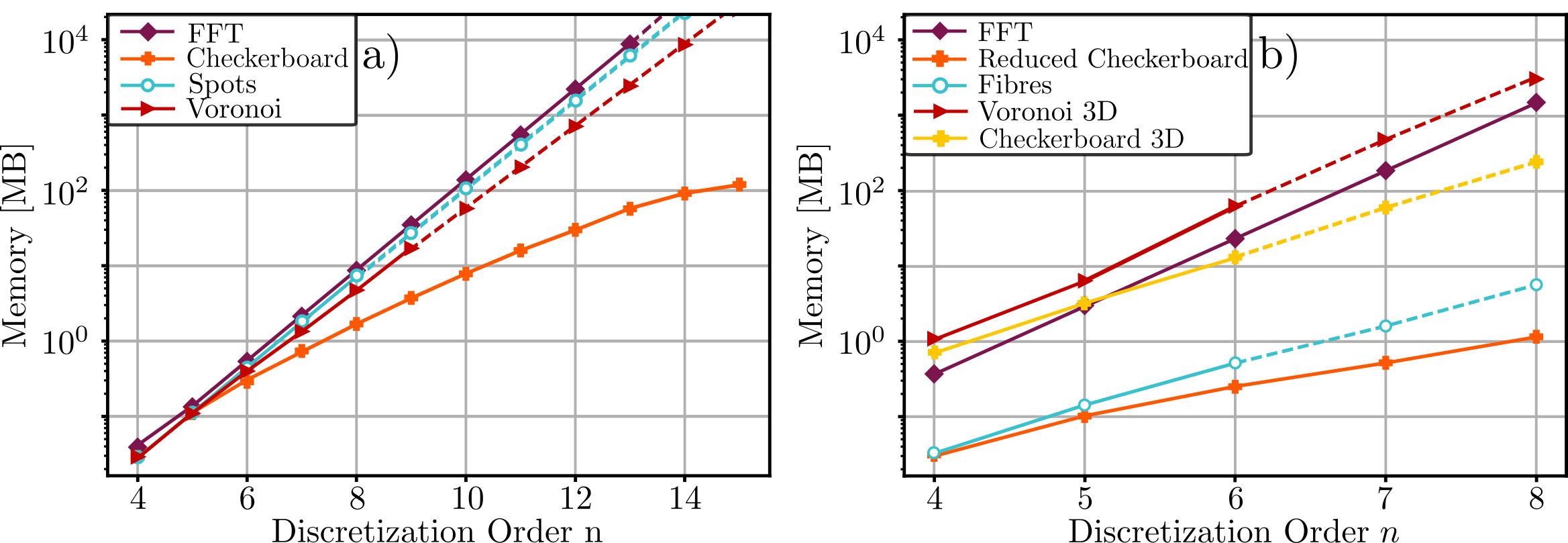}
    \caption{Maximal Memory used for storing the local strain in dependence of the discretization order $n$ in a) 2D
and b) 3D space.}
\label{fig:appendix_memory}
\end{figure}

\begin{figure}[ht]
    \centering
    \includegraphics[width=\textwidth]{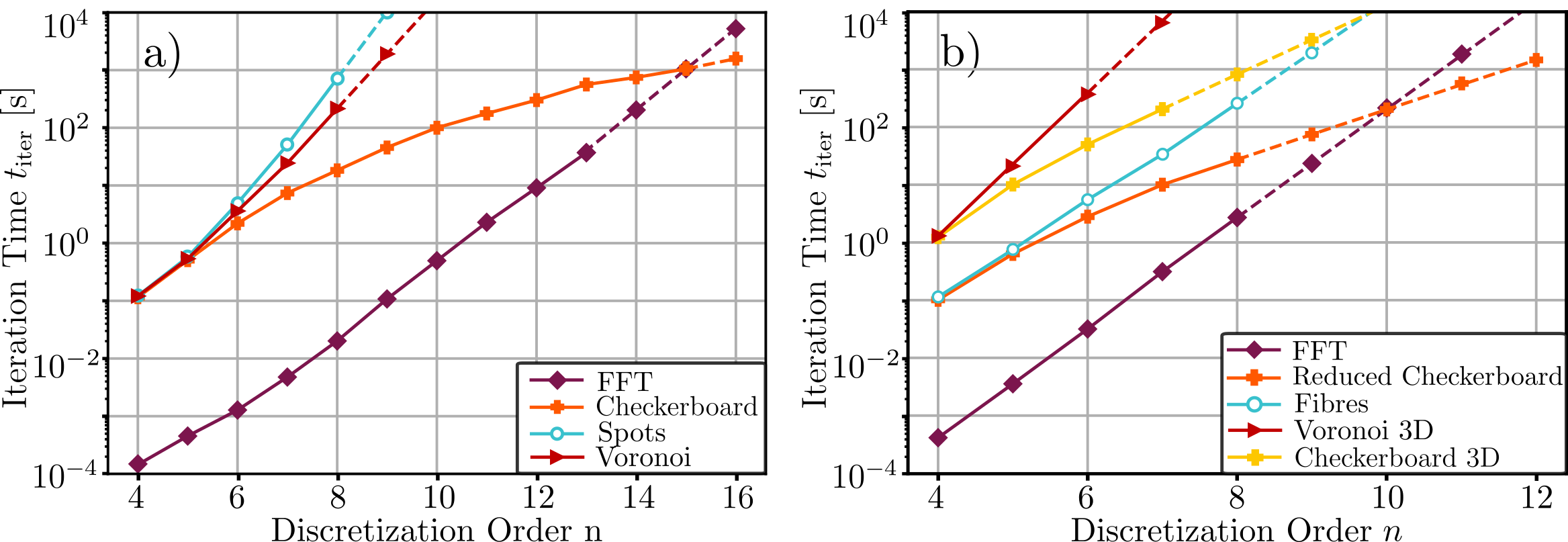}
    \caption{Average iteration time $t_\mathrm{iter}$ in dependence of the discretization order $n$ in a) 2D and b) 3D space.}
    \label{fig:appendix_time}
\end{figure}

\begin{figure}[ht]
    \centering
    \includegraphics[width=\textwidth]{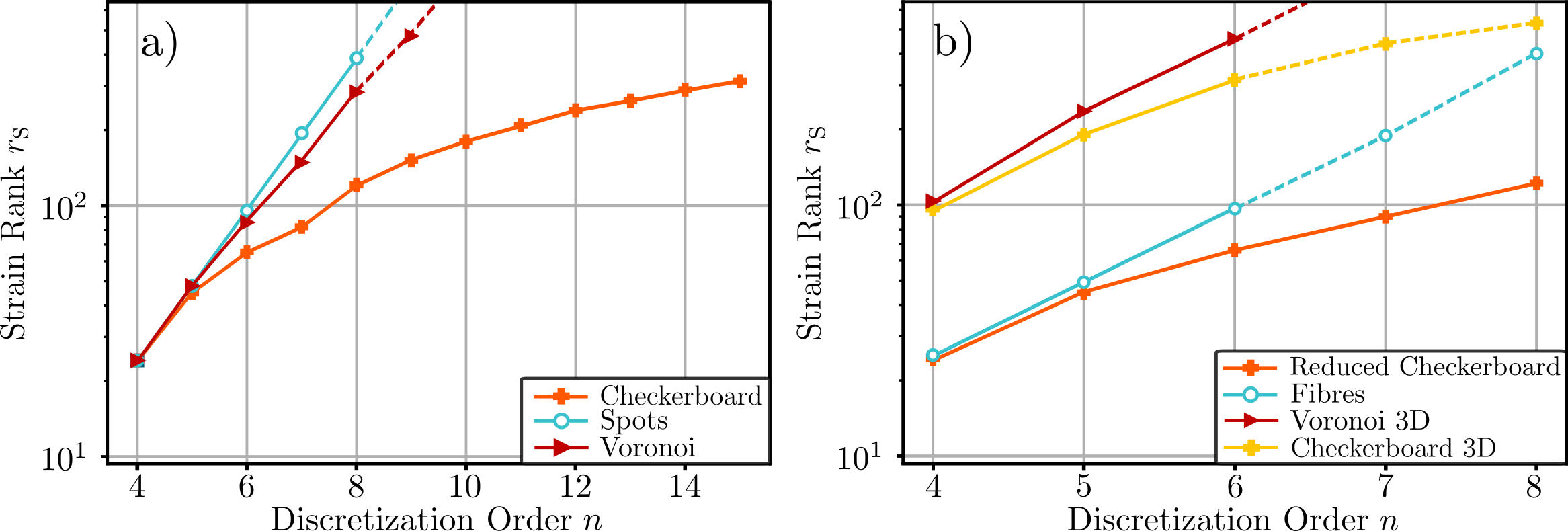}
        \caption{Maximal Strain Rank $r_\mathrm{S}$ during the evaluation of the algorithm in dependence of the discretization order $n$ in a) 2D and b) 3D space.}
    \label{fig:appendix_rank}
\end{figure}

\begin{figure}[ht]
    \centering
    \includegraphics[width=\textwidth]{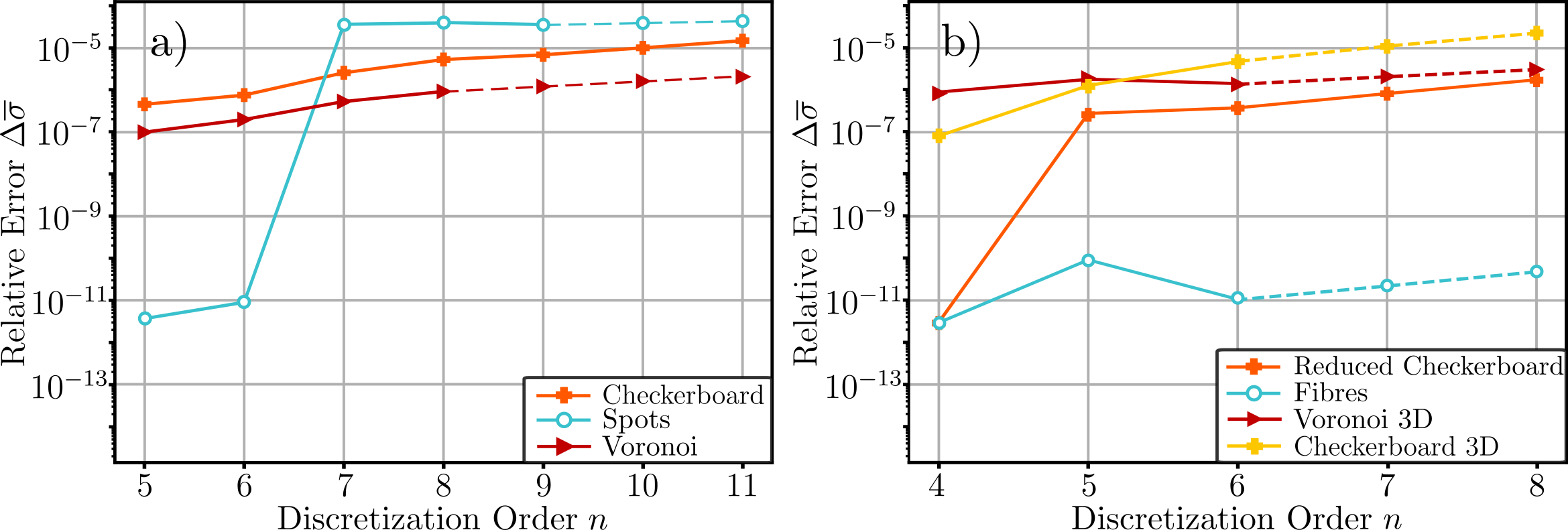}
    \caption{Relative Error $\Delta\overline{\sigma}$ in dependence of the discretization order $n$ in a) 2D and b) 3D space.}
    \label{fig:appendix_error}
\end{figure}

The Spots and Voronoi geometries were generated by sampling a fixed number of points from a uniform distribution over the RVE, which serve as the centers for the inclusions. If any Spots or Voronoi regions overlap, a new set of points is sampled. To ensure consistency across experiments, the random number generator's seed was fixed

For the Voronoi geometry,
a large set of points was initially sampled,
followed by the construction of their corresponding Voronoi regions.
The final inclusions were obtained by randomly selecting a fraction of these points,
in accordance with the predefined volume fraction $\Phi_{vol} = 0.9$.
The material parameters for the selected Voronoi regions were then assigned to the inclusion parameters,
as shown in Figure \ref{fig:appendix_geometries}.
For a more detailed derivation on the construction of the Voronoi geometry,
we refer the interested reader to \cite{sato24}.

The memory usage results,
presented in Figure \ref{fig:appendix_memory},
demonstrate the stark contrast between the exponential scaling of the FFT-based method and the geometry-optimized efficiency of the SFFT-based approach in both 2D and 3D.
For the Checkerboard and reduced Checkerboard geometries,
the SFFT method significantly reduces memory consumption.
On the other hand,
the Voronoi and Voronoi 3D geometries exhibit exponential memory growth, 
similar to the FFT-based method.
For generalized geometries like the Checkerboard 3D,
the SFFT method still offers superior memory efficiency compared to the FFT baseline.
More complex geometries like the Spots,
Voronoi,
Fibres or Voronoi 3D geometries all show exponential memory scaling similar to the state-of-the-art FFT based Method.
However,
due to the redundant z-component of the Fibres geometry, we still see a considerable memory reduction in comparison to the FFT-based method.

For the iteration times,
as shown in Figure \ref{fig:appendix_time},
we observe that the FFT-based method is faster for lower discretization orders.
However,
for all the from the Checkerboard derived geometries,
the SFFT-based method starts to outperform FFT-based Homogenization algorithm after reaching a geometry-dependent crossover point.
In 2D, the crossover occurs at $n=15$ for the Checkerboard geometry,
while in 3D,
the reduced Checkerboard geometry shows a crossover around 
$n=10$.
For the full 3D Checkerboard geometry,
the crossover is expected near 
$n=15$,
however,
this extrapolation extends too far beyond the tested range to be considered reliable.

In Figure \ref{fig:appendix_rank},
we see that the rank growth of the strain TT for the Checkerboard and its two generalizations exhibit sub-exponential scaling,
while Voronoi,
Spots and their 3D counterparts exhibit exponential rank growth. 
This is consistent with the previous results on iteration time and memory consumption.

The dependants of the relative error on the discretization order $n$ can be seen in Figure \ref{fig:appendix_error}. Here, all geometries reach accuracies of $10^{-5}$ or better. 
Thus,
they follow the same trend as the originally considered geometries in Section 5.

In conclusion, these additional geometries further emphasize that the current SFFT-based Homogenization algorithm is particularly well-suited for rectangular geometries.
However,
the results also highlight that geometries incompatible with the existing tensor network structure exhibit scalability performance similar to that of the traditional FFT-based method or worse.

\section{Discretization Scheme}
\label{app:discretization_scheme}

Before investigating the various geometries, it is essential to determine the most appropriate discretization scheme by analysing the Green-Eshelby Operator. This decision is not straightforward, as the iteration time depends on multiple factors: the smoothness of the solution field (and consequently its rank), as well as the rank of the Green-Eshelby Operator itself. While more complex discretization schemes may result in smoother solution fields, they could also lead to a higher rank in the Green-Eshelby Operator, potentially impacting the efficiency of the algorithm.

To assess this, different discretization schemes - listed in Table \ref{tab:discretization_schemes} - were tested for the SFFT-based Homogenization algorithm using a Checkerboard geometry in 2D space. The rounding accuracy was set to a threshold of $\delta_\mathrm{acc} = 10^{-8}$.

\begin{figure}[ht]
    \centering
    \includegraphics[width=\linewidth]{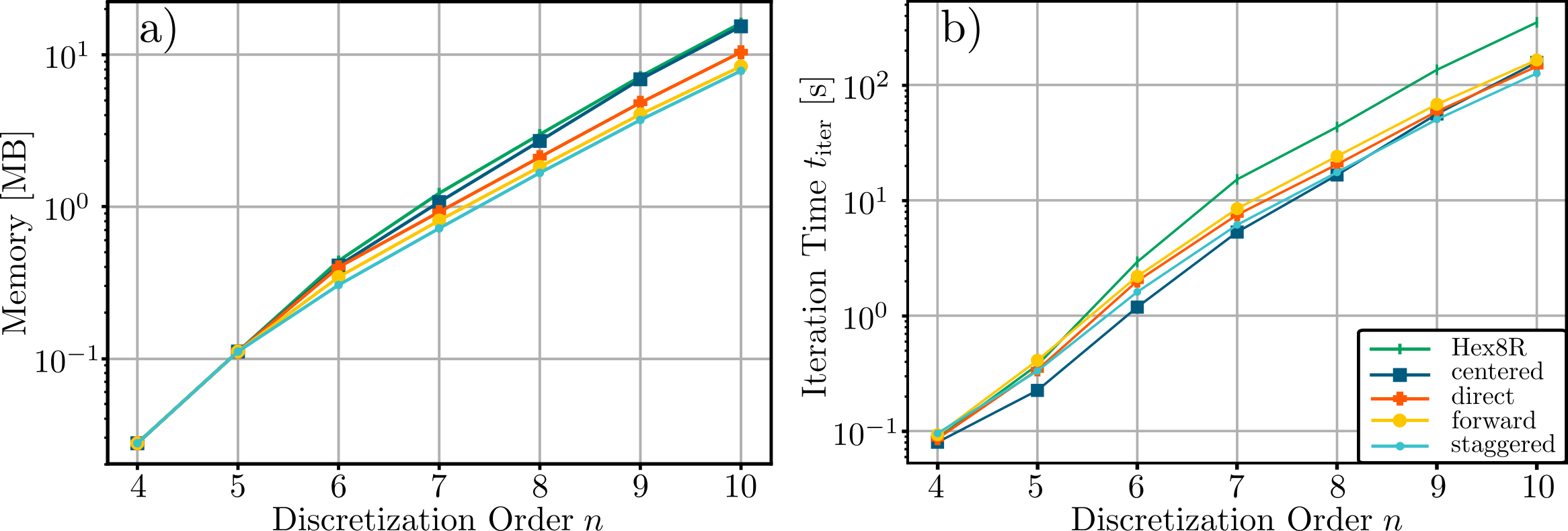}
\caption{a) Maximal memory used for storing the local strain and b) average iteration time $t_\mathrm{iter}$ in dependence of the discretization order $n$.}
    \label{fig:scheme_memory_iter}
\end{figure}

\begin{figure}[ht]
    \centering
    \includegraphics[width=\linewidth]{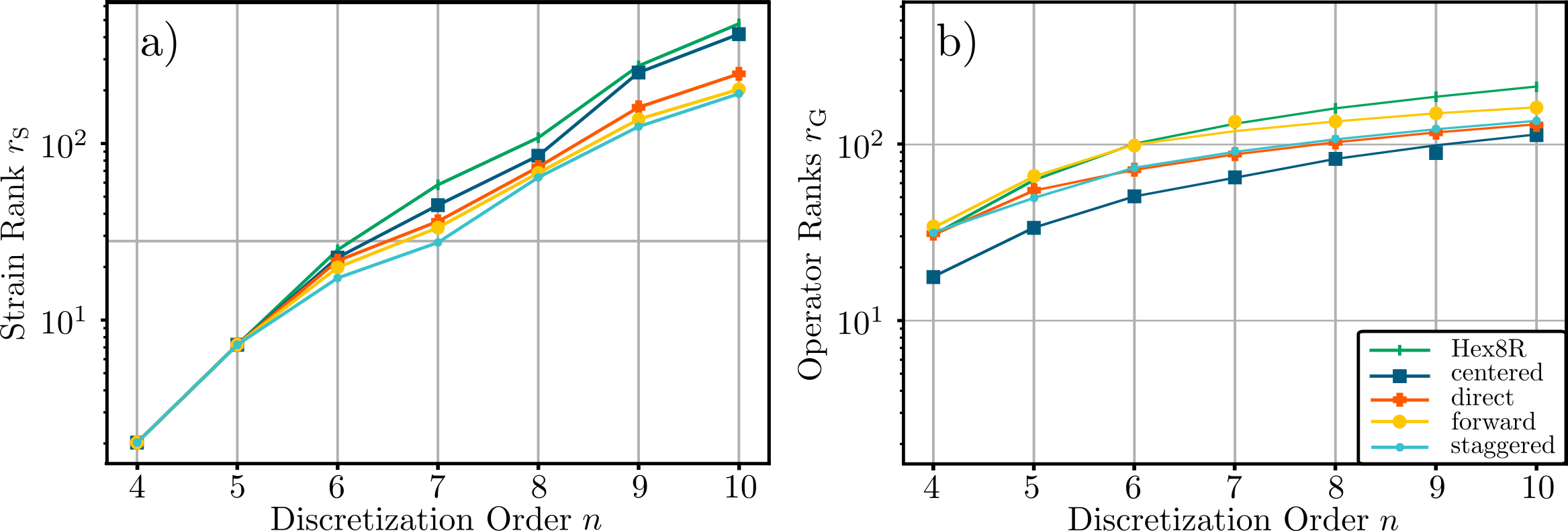}
\caption{a) Maximal local strain rank $r_\mathrm{S}$ during the evaluation of the algorithm and b) Green-Eshelby operator rank $r_\mathrm{G}$ in dependence of the discretization order $n$.}
    \label{fig:scheme_rank}
\end{figure}

As shown in Figure \ref{fig:scheme_memory_iter}, the staggered grid discretization outperforms all other schemes in terms of both memory usage and computational speed.
However, the speed of the centered grid scheme dominates at lower discretization orders but is eventually surpassed by the staggered grid at higher orders. This result is closely tied to the strain ranks, as depicted in Figure \ref{fig:scheme_rank} a), where the staggered grid shows lower ranks, indicating a smoother solution field.

Figure \ref{fig:scheme_rank} b) shows that the Green-Eshelby Operator in the centered grid scheme has a lower rank than in the staggered grid for the considered discretization range. However, at higher discretization orders, the staggered grid surpasses the centered grid due to its improved scaling behaviour.

The iteration times and strain rank growth in the centered scheme reveal an interesting phenomenon: hourglassing. This numerical artifact introduces spurious, low-frequency oscillations that do not correspond to real physical deformations, increasing the strain rank. As a result, iteration times rise despite the centered grid having the lowest Green-Eshelby operator rank.

\end{document}